\documentclass[12pt, draftclsnofoot, onecolumn]{IEEEtran}

\usepackage[pdftex]{graphicx}
\usepackage[cmex10]{amsmath,empheq}
\usepackage{cite}
\usepackage{url}
\usepackage{color}
\usepackage{microtype}
\usepackage{amssymb}
\usepackage{relsize}
\usepackage{lipsum}
\usepackage{epstopdf}
\usepackage{enumitem}
\usepackage{bm}
\usepackage{nicefrac}
\usepackage[percent]{overpic}
\usepackage{subfig}

\UseRawInputEncoding

\def\R{\mathbb{R}}
\def\C{\mathbb{C}}

\def\NT{N_{\mathrm T}}
\def\NR{N_{\mathrm R}}
\def\Nb{N_{\mathrm b}}

\begin{document}

\title{DeepTx: Deep Learning Beamforming with Channel Prediction}

\newcommand{\changed}[1]{{\color{blue}{#1}}}
\newcommand{\todo}[1]{{\color{red}{#1}}\PackageWarning{TODO:}{#1!}}

\author{Janne~M.J. Huttunen,
        Dani~Korpi
        and Mikko~Honkala
~
\thanks{J.M.J. Huttunen, D. Korpi and M. Honkala  are with Nokia Bell Labs, Espoo, Finland.}
}

\maketitle

\begin{abstract}

  Machine learning algorithms have recently been considered for many tasks in the field of wireless communications. Previously, we have proposed the use of a deep fully convolutional neural network (CNN) for receiver processing and shown it to provide considerable performance gains. In this study, we focus on machine learning algorithms for the transmitter.  In particular, we consider beamforming and propose a CNN which, for a given uplink channel estimate as input, outputs downlink channel information to be used for beamforming. The CNN is trained in a supervised manner considering both uplink and downlink transmissions with a loss function that is based on UE receiver performance. The main task of the neural network is to predict the channel evolution between uplink and downlink slots, but it can also learn to handle inefficiencies and errors in the whole chain, including the actual beamforming phase. The provided numerical experiments demonstrate the improved beamforming performance.
\end{abstract}


\section{Introduction}

Machine learning (ML) has recently been considered as a potential tool to improve the performance of next-generation wireless systems; see for example \cite{Zhang19a,Jiang17a,Fu18a}. Especially, there has been growing interest in implementing radio physical layer algorithms with ML \cite{oshea17,huang20,zhao2018,neumann18,chen19,gunduz19}. In recent works, ML has been applied to improve the detection accuracy of a physical-layer receiver \cite{honkala21,korpi21,pihlajasalo21}, learn new and better waveforms \cite{korpi2022waveform,goutay2021learning,aoudia2021endtoend}, as well as reduce the complexity of specific parts of receiver algorithms \cite{shental2019}. In addition, ML introduces other benefits, including efficient inference using ML accelerators and reduced amount of manual labor as the algorithms are automatically ``programmed''  using data. ML can also improve performance by facilitating adaptation to specific environments via re-training or fine-tuning.

In this paper, we continue on the path of implementing radio physical layer algorithms with ML. Building on our earlier work on ML-based receiver algorithms \cite{honkala21,korpi21,pihlajasalo21}, we consider now the radio transmitter by applying convolutional neural networks (CNNs) to transmit beamforming or precoding in time-division duplexing (TDD), multiple-input and multiple-output (MIMO) radio systems. The main benefit of the proposed solution is the reduction of the harmful effects of channel aging at the base station (BS), which is an inherent issue in TDD systems. This means that the CNN-based beamformer can operate with older channel information, allowing for sparser reference signals for channel estimation and/or increased flexibility in the scheduling of uplink (UL) and downlink (DL) traffic.

\subsection{Related work and motivation}
\label{sec:relat-work-motiv}

Current radio systems typically rely on techniques such as eigenbeamforming and zero forcing (ZF) when calculating the beamforming coefficients. More advanced algorithms are also developed (see e.g. \cite{Bjornson2014}), but those are typically also more computationally demanding. Recently, there have also been several studies considering application of ML in beamforming. For example, \cite{zhang21,kaya21} propose ML solutions for selection of pre-computed beams.  In addition, \cite{Xia2020,Shi2020} propose an approach  in which the channel state information is input to a CNN network to produce a prediction of key parameters in traditional beamforming algorithms. In addition, \cite{huang19} proposes an unsupervised learning solution for beamforming.  For frequency-division duplex (FDD) systems, \cite{Kong2021} proposes the use of ML for predicting a code book index (an integer) which is transmitted from UE to BS for calculation of beamforming coefficients.

Most of these techniques, however, treat the obtaining of the channel estimates to be used by the beamformer as a separate problem. In 5G TDD systems, for instance, the source of this channel information is often the sounding reference signal (SRS) transmitted by the UEs, which can be used to estimate the reciprocal DL channel, this estimate being then used by the beamformer to calculate the precoding coefficients. However, this type of approach is accurate only for slowly fading channels or if the channel does not significantly change between the UL and DL slots. This rarely holds for mobile UEs, especially if they are moving with high velocity. It has been shown that such channel aging can significantly degrade the radio performance, especially in the case of DL beamforming \cite{hu18,isukapalli09,kim14}.

There are a variety of methods that aim at predicting channel evolution. For example, channel prediction can be carried out by parametrizing the channel model and applying autoregressive models (e.g. \cite{eyceoz98,arredondo02}) or Kalman filter (e.g. \cite{sternad03,jiang19}) to the channel parameters. However, in this case the prediction method is only working if the chosen parameterized channel model is valid for the environment.

ML is also widely applied to channel estimation and prediction. For example, \cite{ding14} applies complex neural networks to channel prediction. In \cite{dong19}, on the other hand, CNN-based approach for estimating the channel from pilot information is proposed.  There, LSTMs are applied for predicting the channel over a transmission time interval (TTI). Various recurrent neural network (RNN)-based approaches are proposed in \cite{potter08,routraya12,jiang18,jiang19,jiang20,jiang20b}, which consider channel prediction over a single frequency. The work in \cite{alrabeiah19} proposes using a deep fully connected neural network to map a channel from a set of antennas and frequencies to another set of antennas and frequencies possibly in a different location. The proposed method is applicable for channel prediction between UL and DL in FDD and distributed MIMO TDD scenarios. However, time evolution of channels is not considered in \cite{alrabeiah19}. Furthermore, channel prediction methods combining CNNs and RNNs are proposed in \cite{liao19,luo20}, while a CNN-based approach for estimating the channel aging pattern is proposed in \cite{yuan19}. Channel prediction for FDD massive MIMO systems is proposed in \cite{arnold2019,wang19,alrabeiah19,yang2020,mashhadi21}. These studies, however, focus solely on channel information and do not as such consider beamforming.

\subsection{Contributions and Organization}
\label{sec:contr-organ}

As opposed to prior art, in this paper we consider \emph{the joint task of channel prediction and beamforming} in TDD systems. This means that the proposed approach does not separate the channel estimation from the beamforming, instead utilizing ML to optimize the complete system by maximizing the chosen key performance indicator based on UE performance. In particular, instead of using model-based parametrization or reinforcement learning, we propose a \textit{supervised} approach for beamforming that applies a CNN to predict evolution of channel between UL and DL. The CNN takes UL channel estimate as input and outputs channel data that is fed to DL beamformer. Crucially, the optimization of this CNN is done based on the final detection accuracy at the UE receivers, meaning that also the effect of the beamformer is incorporated into the overall model, although only the channel predictor CNN is being trained. 

In particular, the main contributions are as follows: 
\begin{itemize}
  \item We propose a novel deep learning transmitter (DeepTx) based on residual neural networks which predicts sufficient channel information for beamforming. In addition to the prediction of the future channel, DeepTx also learns to pre-augment the input to the ZF beamformer to handle inefficiencies and errors in the ZF beamforming, thereby improving performance.
	\item Both UL and DL and the evolution of channel are considered, meaning that channel information is obtained from a realistic UL channel estimate and applied for DL beamforming  as in practical beamforming setups. 
  \item We design a supervised learning approach in which the loss function is directly based on UE receiver performance (uncoded bits). This means that both in DL and UL, the transmitters, channels and receivers needs to be considered and implemented in a differentiable manner so that one can backpropagate through the whole chain. The benefit of this approach is that it can learn such CNN weights that maximize the overall system performance instead of some intermediate metric, such as channel estimate accuracy. In other words, CNN can focus only on those aspects of the channel that are meaningful for the final performance.
	\item We provide an extensive set of numerical results based on 4x2 MIMO scenarios. The obtained results show that DeepTx can outperform a conventional channel estimator and beamformer by a large margin. 
        \end{itemize}

Although in this paper ZF is chosen as the beamformer, the proposed approach is generic in nature and it could be replaced with any  differentiable beamformer. On the other hand, ZF can also be seen as an additional non-linear layer in the neural network. Namely, in our previous work with a neural network receiver, standard residual neural networks were found out to work in simpler scenarios and expert knowledge based nonlinear layers were required to reach good performance in more complex MIMO \cite{korpi21}. In this current work, ZF can be seen as similar expert knowledge based layer. In principle, ZF can also be disregarded in which case DeepTx would output precoding matrix or beamforming output directly, but  setup was found out to be insufficient indicating that some expert knowledge based nonlinearity is also needed in this transmitter case.

The rest of this paper is organized as follows. In Section~\ref{sec:simulator}, we describe the system model including models for uplink and downlink transmissions. Traditional ZF beamforming is also explained. Section \ref{sec:nn_tx} describes in detail the proposed CNN based approach for beamforming.    The proposed approach is validated in Section~\ref{sec:validation_res} using numerical results. Conclusions are given in Section~\ref{sec:conc}.

\section{System Model} 
\label{sec:simulator}

\begin{figure}[ht!]
	\centering
	\includegraphics[width=0.5\linewidth]{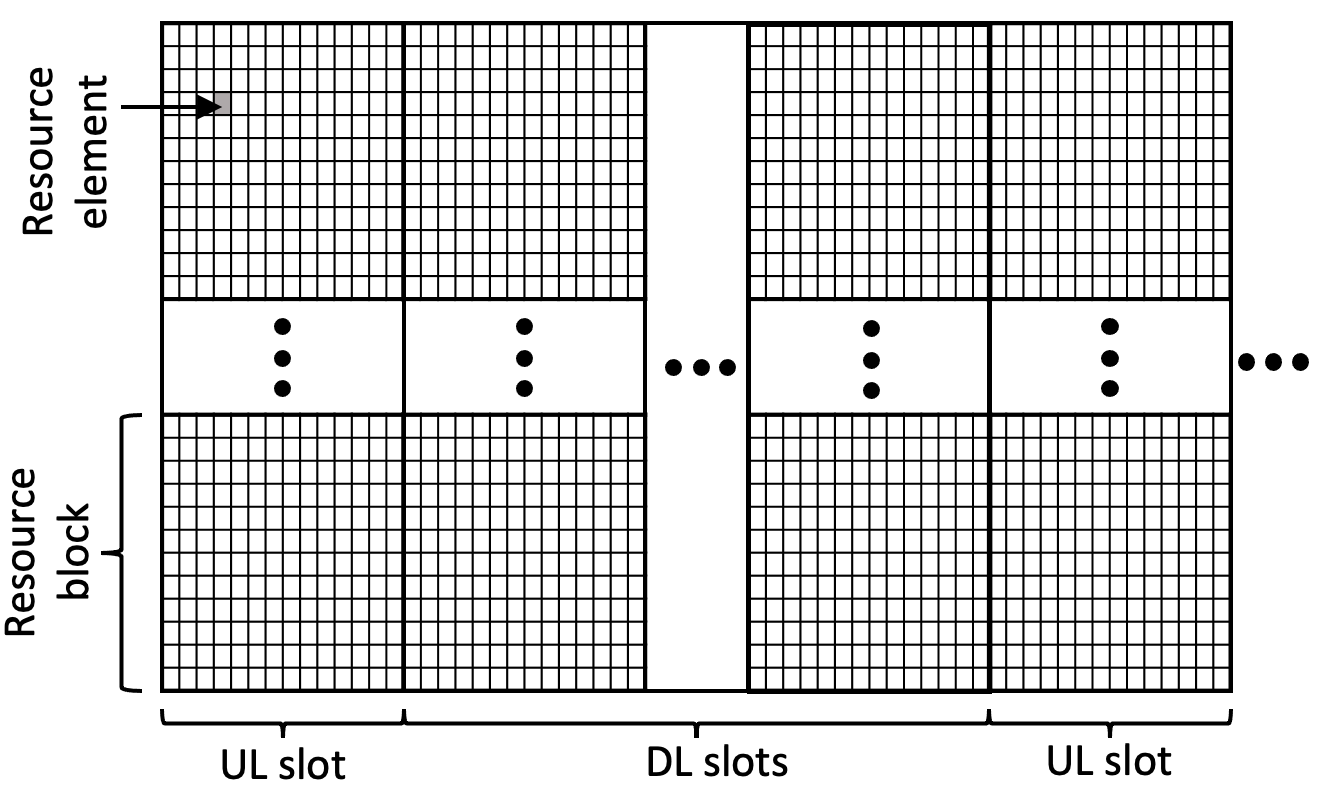}
	\caption{Time division duplexing presented in frequency domain. For example, in 5G, UL and DL slots include 14 OFDM symbols and a given number of resource blocks, each resource block including 12 subcarriers. }
	\label{fig:tddslots}
\end{figure}

We consider TDD system in which UL and DL transmissions are carried out in different time slots (see Fig. \ref{fig:tddslots}), but on the same frequency band. The base station is equipped with $\NR$ antennas and is capable of MIMO transmission with $\NT$ layers. Our analysis covers two scenarios: one in which there is a single UE with $\NT$ antennas (single user or SU-scenario) and another where there are $\NT$ UEs equipped with only one antenna each (multi user or MU-scenario). We assume OFDM transmission with $S$ symbols (typically 14) and $F$ subcarriers. 

\subsection{Uplink Signal Model}

Let $\mathbf s_{ij}\in \C^{\NT}$ be the vector of the symbols transmitted in $ij$th resource element (RE, $i=1,\ldots,S$ and $j=1,\ldots,F$). The signal received at the base station $\mathbf y_{ij}\in\C^{\NR}$ can be expressed as \begin{displaymath}
  \mathbf {y}_{ij}=\mathbf {H}_{ij}^\textrm{UL}\mathbf {s}_{ij}+\mathbf{n}_{ij}
\end{displaymath}
where $\mathbf H_{ij}^\textrm{UL}\in\C^{\NR\times \NT}$ is the UL channel matrix and $\mathbf{n}_{ij}\in\C^{\NR}$ is the noise term including observation noise and interference. Note that we assume a system model without inter-symbol interference (ISI) or inter-carrier interference (ICI), which means that the channel can be modeled in the frequency-domain with a single tap per subcarrier. Extending this work to cover also scenarios with ISI and ICI is an important future work item for us.

\subsection{Downlink Signal Model}

Again, let $\mathbf s_{ij}\in\mathcal \C^{\NT}$ be the vector of the transmit symbols. In DL, we apply precoding (or beamforming), in which case the output signal from the antennas can be expressed as
\begin{displaymath}
  \mathbf x_{ij}=\mathbf W_{ij}\mathbf s_{ij}
\end{displaymath}
where $\mathbf W_{ij}\in\C^{\NR\times \NT}$ is a precoding matrix. The received signals by the UEs (all signals stacked to a vector $\mathbf y$) can be written as
\begin{displaymath}
  \mathbf y_{ij}=\mathbf H_{ij}^\textrm{DL}\mathbf x_{ij}+\mathbf n'_{ij}=\mathbf H_{ij}^\textrm{DL}\mathbf W_{ij}\mathbf s_{ij}+\mathbf n'_{ij}
\end{displaymath} where $\mathbf H_{ij}^\textrm{DL}\in\C^{\NT\times \NR}$ is the DL channel matrix, and $\mathbf n'_{ij}$ is the noise signal.

Due to the channel reciprocity, the channel to DL direction is assumed to be same as the UL direction: $\mathbf H_{ij}^\textrm{DL}$ is the transpose of $\mathbf H_{ij}^\textrm{UL}$. Strictly speaking, this is a valid assumption for time-variant channels only when UL and DL transmissions happen at the same time (we discuss this later). In the following, we simplify notation by omitting the superscript UL/DL from $\mathbf H$ when confusions can be avoided.

\subsection{Conventional Receiver Processing}

We review shortly the basic components of receiver processing, which are applied both in the UL and DL receivers (UE and BS, respectively) unless mentioned otherwise. Moreover, the proposed approach relies on an channel estimate in UL, while DL receiver processing in UEs is utilized in the loss function.  

\subsubsection{Channel estimation} The channel is estimated using reference signals, also referred to as pilots. For UL transmission, the channel is typically estimated using demodulation reference signals (DMRS). However, as DMRS reference signals are only scheduled on those physical resource blocks (PRBs) which are allocated to the user for data transmission and the allocation can change between UL and DL, the channel for DL beamforming is typically estimated using SRSs that cover the whole bandwidth. Assuming that the pilot symbols lie on the unit circle, such as when using quadrature phase shift keying (QPSK) pilots, the channel estimate is calculated with least squares (LS) as follows: 
\begin{displaymath}
  \widehat{\mathbf H}_{ij} = \mathbf y_{ij} ^* \mathbf x_{ij},\quad (i,p)\in\mathcal P,
\end{displaymath}
where  $(\cdot)^*$ denotes the Hermitian transpose, $\mathcal P$ denotes the indices of REs carrying pilots and $\mathbf x_{ij}$ and $y_{ij}$ are the transmitted and received pilot symbols. The channel estimate is expanded to the whole slot using linear interpolation.

\subsubsection{Equalization and demodulation}

After obtaining the interpolated channel estimate, each data symbol is equalized using linear minimum mean squared error (LMMSE) equalizer. The post-equalized data symbols are obtained as
$ \hat{\mathbf x}_{ij} = \mathbf D_{ij} \mathbf V_{ij} \mathbf{y}_{ij}$, 
where the LMMSE equalizer is given by
\begin{displaymath}
  \mathbf V_{ij} = \left(\widehat{\mathbf{H}}_{ij}^* \widehat{\mathbf{H}}_{ij} + \hat{\sigma}^2_n \mathbf{I}\right)^{-1} \widehat{\mathbf{H}}_{ij}^* ,
\end{displaymath}
where $\hat{\sigma}^2_n$ is the noise power estimate, $\mathbf{I}$ is the identity matrix and $\mathbf D_{ij}$ is the inverse of the diagonal part of the matrix $\mathbf V_{ij}\widehat{\mathbf{H}}_{ij}$, the purpose of which is to re-scale the LMMSE equalizer output\cite{goutay21}.

Assuming that noise of the equalized symbols is Gaussian, LLRs of $m$th bit transmitted by the $k$th user in RE ($i$,$j$) can be expressed as
\begin{equation}
  \label{eq:1}
  \mathrm{LLR}_{ijk}(m)=\log\left(\frac
      {\sum_{c\in\mathcal C_{m,0}}\exp\left(-\frac1{\sigma_{ijk}^2}\left\vert \hat{\mathbf x}_{ijk}-c\right\vert^2\right)}
      {\sum_{c\in\mathcal C_{m,1}}\exp\left(-\frac1{\sigma_{ijk}^2}\left\vert \hat{\mathbf x}_{ijk}-c\right\vert^2\right)}
  \right)
\end{equation}
where $\mathcal C_{m,0/1}$ is the subset of the constellation points with $m$th bit set to 0/1 and $\sigma_{ijk}^2$ is the estimated noise variance. 

\subsection{Zero-Forcing Beamforming}
\label{sec:trad-rece-proc}

In this paper we consider only ZF beamforming which is shortly described here. Other techniques for specifying the precoding matrix $\mathbf W$ can be found, for example, in \cite{Bjornson2014}.

Objective of beamforming is to direct the signals to those UEs which are meant to receive it, while avoiding mutual interference between the signals or caused to other UEs. Restricting the analysis to a single OFDM subcarrier which experiences frequency-flat fading, such beamforming results in $\mathbf y_{ij}=\mathbf s_{ij}+ \mathbf {\tilde n}_{ij}$, i.e., the receivers will receive their intended signal plus some (uncorrelated) noise. One method for achieving this is to choose $\mathbf W_{ij}$ such that $\mathbf H_{ij}\mathbf W$ becomes an identity matrix. This can be achieved if $\mathbf W_{ij}$ is chosen to be the pseudoinverse of $\mathbf H_{ij}$,
\begin{equation}\label{eq:ZF}
  \mathbf W_{ij}= \mathbf H_{ij}^\dagger=\mathbf H_{ij}^* (\mathbf H_{ij}\mathbf H_{ij}^*)^{-1}.
\end{equation}
This choice  is referred to as the ZF beamformer. 

The main challenge of accurate beamforming is obtaining the required channel information ($\mathbf H_{ij}$). In TDD networks, the reciprocity of the channel can be utilized such that the DL channel is approximated using UL channel estimate. This approximation can be accurate if the channel does not change significantly between UL and DL time slots.  However, for fast-fading channels, the channel estimate obtained during UL can already be outdated and beamforming carried out using UL channel estimate can lead to significant performance degradation \cite{hu18,isukapalli09,kim14}, despite the equalization that is performed in the receiver.  To tackle this problem, we apply CNNs to predict the channel evolution. The CNN can observe the variability of the channel in training data and utilize this information to handle such variability in the prediction.

\section{Convolutional Neural Network-Based Beamformer}
\label{sec:nn_tx} 

In this work, we design a CNN-based beamforming solution, which aims at predicting channel evolution between UL and DL slots. The proposed approach, referred to as DeepTx, is based on a ResNet type neural networks \cite{he2016identity}, which takes the UL channel estimate as input and processes it to output the precoding matrices.

\begin{figure}[ht!]
	\centering
	\includegraphics[trim=100px 70px 100px 135px, clip, width=1\linewidth]{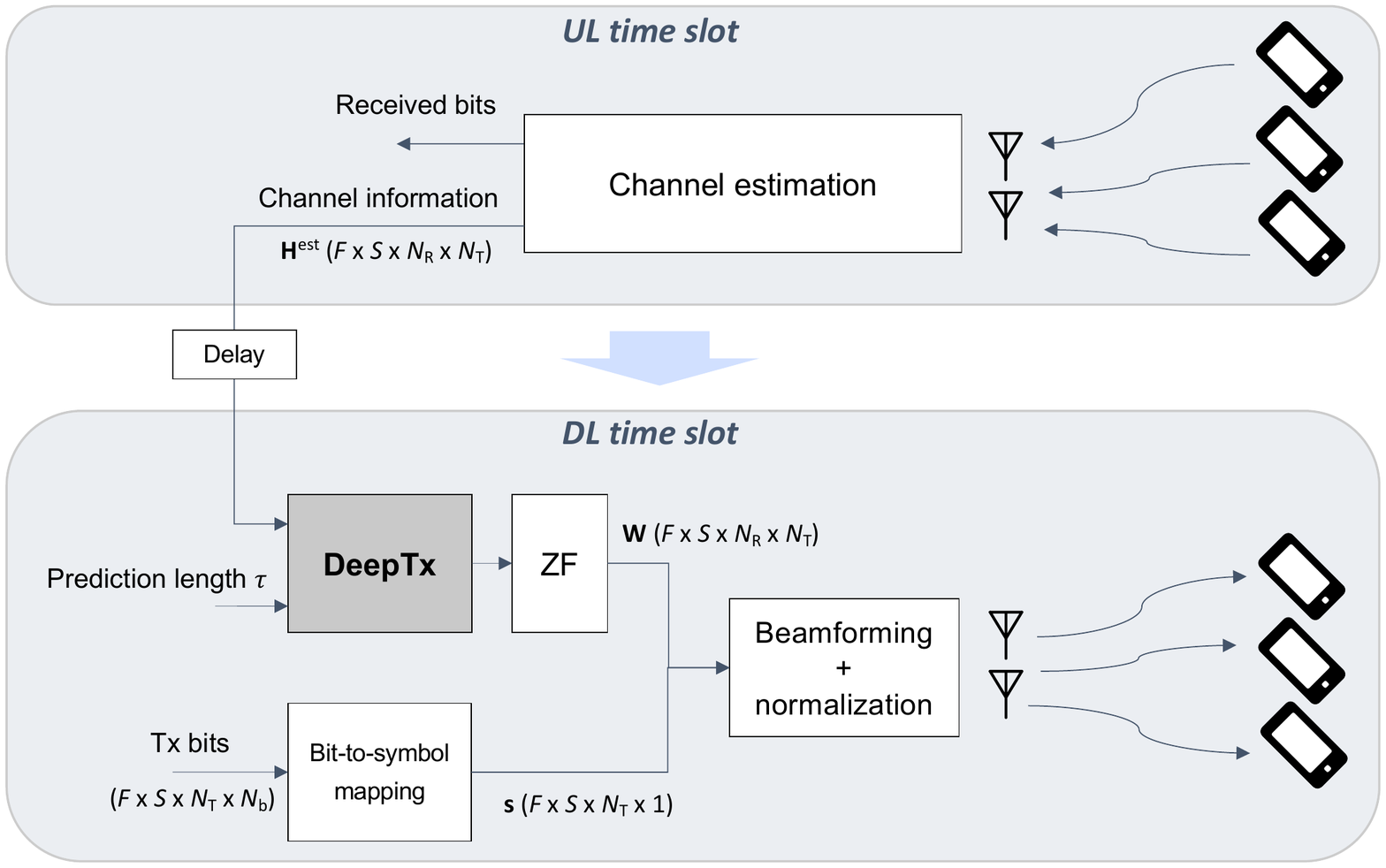}
	\caption{An example of the operation of DeepTx with three UEs. $F$ and $S$ are the number of subcarriers and symbols in a slot, $\NR$ is the number antennas in BS and $\NT$ is the number of MIMO layers, $\Nb$ is the number of bits per symbol. }
	\label{fig:approachA}
\end{figure}

The procedure for DeepTx is described in the following (see Fig.~\ref{fig:approachA} for illustration):
\begin{enumerate}
\item\label{item:approachA_UL} During a UL slot, the receiver processes the received data containing the DMRS to form the LS channel estimate.  This is then interpolated to cover all REs to form a data array $\mathbf {H}^\mathrm{est}$ consisting of $F\times S\times \NT \times \NR$ elements. Note that channel estimation is a typical part of UL processing, i.e., this step does not consume any additional computational resources.
\item\label{item:approachA_DeepTx} The channel estimate $\mathbf {H}^\mathrm{est}$ is transferred to the transmitter and given to DeepTx. The prediction length $\tau$ (the number of slots between UL and DL) is also input to DeepTx. DeepTx gives the prediction of DL channel $\mathbf {H}^\mathrm{pred}$ as output.
\item\label{item:approachA_ZF} The predicted channels $\mathbf {H}_{ij}^\mathrm{pred}$ are transformed to the precoding matrices $\mathbf W_{ij}$ by applying ZF transformation (Eq. \eqref{eq:ZF}). 
\item\label{item:approachA_precoding} Precoded symbols are computed as $\mathbf x_{ij}=\mathbf W_{ij}\mathbf s_{ij}$, where $\mathbf s_{ij}$ are the symbols to be transmitted (possibly including DL DMRS symbols).
\item\label{item:approachA_norm} Power normalization is applied to form signal to be transmitted. For example, the symbols can be scaled as $\mathbf x_{\rm transmitted}=a \mathbf x/\Vert \mathbf x\Vert_{L^2}$ where $\Vert \mathbf \cdot \Vert_{L^2}$ is the $L^2$-norm and $a$ is a constant which scales the TX signal $\mathbf x$ to the desired transmit power. 
\end{enumerate}

{\bf Remark:} The main task of DeepTx is to predict how the channel is evolving between UL and DL slots and provide the input to ZF precoding (pre-corrected to handle possible errors in the ZF approach). One may wonder if the CNN could do more, for example, by producing the precoding matrices $\mathbf W_{ij}$ directly or, even further, producing the actual transmit symbols $\mathbf x_{ij}$. We have also experimented with these setups, but it turned out that the above approach including the explicit ZF transformation and multiplication of $\mathbf x_{ij}$ gives by far the highest performance. The reason might be the high nonlinearity of the precoding, which is difficult to learn for a CNN. The evolution of the channel has more linear behavior and is thereby easier for a convolutional neural network to approximate. One way to see this is that ZF (or, more specifically, \eqref{eq:ZF}) forms a non-linear layer within the neural network which brings expert knowledge into the architecture. A similar phenomenon has been observed with ML-based receivers, where a regular CNN performs worse than an augmented CNN that can also express multiplications \cite{korpi21}. 

\begin{table}[!t]
	\renewcommand{\arraystretch}{1.3}
	\caption{The DeepTx CNN ResNet architecture used in our experiments. The ResNet block is described in \cite{honkala21}.  $L^2$-activation regularization is applied after the output of the network to avoid too large values in the output, which are otherwise not penalized due to the normalization. } 
	\label{tab:cnn_arch}
	\centering
	\footnotesize
	\begin{tabular}{|l|c|c|c|c|}
		\hline
		\textbf{Layer} & \textbf{Type} & \textbf{Filter ($S$, $F$)}& \textbf{Dilation ($S$, $F$)} & \textbf{Output Shape}  \\ \hline\hline
		Input 1 $\mathbf {H}^\mathrm{est}$&  UL channel estimate & & & ($S$, $F$, $\NR$, $\NT$) \\ \hline
          Input 2 $\tau$&  Prediction length (divided by $\tau_\textrm{max}$   & & & ($S$, $F$, $1$) \\ 
           &   and replicated to a $S\times F$ grid)  & & &  \\ \hline
		Input $\mathbf{Z}_c\in\mathbb{C}$&  Reshape Input 1 and concatenate all inputs  & & & ($S$, $F$, $\NR\NT+1$) \\ \hline
		Real input $\mathbf{Z}\in\mathbb{R}$&  $\mathbb{C}\Rightarrow\mathbb{R}$ & & & ($S$, $F$, $2N_c$) \\ \hline
		Conv. In & 2D convolution & (1,1) & (1,1) & ($S$, $F$, 128) \\ \hline
		ResNet Block 1 &  Depthwise separable conv. & (3,3)& (1,1) & ($S$, $F$, 128) \\ \hline
		ResNet Block 2 &  Depthwise separable conv. & (3,3)& (1,1) & ($S$, $F$, 128) \\ \hline
		ResNet Block 3&  Depthwise separable conv.  & (3,3)& (2,3) & ($S$, $F$, 128) \\ \hline
		ResNet Block 4 &  Depthwise separable conv. & (3,3)& (2,3) & ($S$, $F$, 256) \\ \hline
		ResNet Block 5 &  Depthwise separable conv. & (3,3)& (3,5) & ($S$, $F$, 256) \\ \hline
		ResNet Block 6 &  Depthwise separable conv. & (3,3)& (3,5) & ($S$, $F$, 256) \\ \hline
		ResNet Block 7 &  Depthwise separable conv. & (3,3)& (3,5) & ($S$, $F$, 256) \\ \hline
		ResNet Block 8 &  Depthwise separable conv. & (3,3)& (2,3) & ($S$, $F$, 256) \\ \hline
		ResNet Block 9 &  Depthwise separable conv. & (3,3)& (2,3) & ($S$, $F$, 128) \\ \hline
		ResNet Block 10 &  Depthwise separable conv. & (3,3)& (1,1) & ($S$, $F$, 128) \\ \hline
		ResNet Block 11 &  Depthwise separable conv. & (3,3)& (1,1) & ($S$, $F$, 128) \\ \hline
		Conv. Out & 2D convolution & (1,1) & (1,1) & ($S$, $F$, $2\NR\NT$) \\ \hline
		Back to complex &  $\mathbb{\R}\Rightarrow\mathbb{\C}$ & & & ($S$, $F$, $\NR\NT$) \\ \hline
		Reshape  & Output & & & ($S$, $F$, $\NR$, $\NT$) \\ \hline
	\end{tabular}
\end{table}

DeepTx is built with pre-activation ResNet blocks \cite{he2016identity}, and its detailed architecture is described in Table~\ref{tab:cnn_arch}. Convolutions are chosen to be depthwise separable convolutions (DSC) \cite{sifreM13,chollet2017}. DSCs are significantly cheaper to compute compared to normal CNNs and were also experimentally validated to outperform normal CNNs in DeepRx \cite{honkala21}. In our architecture, $\tau$ is normalized to the unit interval and concatenated with the other model inputs. Even though the input to ResNet blocks typically involves some kind of normalization, we have not applied normalization to $\mathbf {H}^\mathrm{est}$. However, we note that the produced channels are normalized during the actual data generation by setting the total power of the path gains, averaged over time, equal to 0~dB. Therefore, the channel estimates have approximately the same range, except for channel estimation errors. Furthermore, we note that this is only one option and other approaches, such as feeding $\tau$ through a bias term of a CNN layer, are also possible and typically lead to similar results, as long as information is allowed to flow through most of the layers. 

\subsection{Generation of data and training procedure }
\label{sec:training_data}

DeepTx is trained based on DL performance by optimizing a loss function which compares the output of UEs (detected uncoded bits) to ground truths (transmitted bits).  As in \cite{honkala21,korpi21,pihlajasalo21}, training is carried out by using simulated data generated using a link level simulator. However, there are a few additional requirements compared to training just a receiver. First, we need to simulate both UL and DL and also the time evolution (or aging) of the channel between UL and DL slots. The effect of the channel also needs to be differentiable with the derivative implemented (e.g. using Tensorflow or other toolbox with automatic derivation). However, it is not necessary to carry out the whole channel simulation during the training procedure. For example, one can precompute time domain (e.g. an impulse response) or frequency domain presentation (e.g. a channel matrix) of the channel offline. In addition, the UE receiver algorithm also needs to be implemented in a differentiable manner. 

In training, one ``forward pass'' involving the sample generation and DeepTx inference can be carried out using the following procedure: 
\begin{enumerate}
\item{}  [CH EVOL] For a randomly chosen $\tau=1,\ldots,\tau_{\max}$, where $\tau$ denotes the number of slots between UL and DL, simulate the evolution of channel from UL slot to DL slot. Store information about the channel (time or frequency presentation).
\item{}  [UL UE-Tx] Simulate UEs with random parameters: generate transmitted UL bits, map bits to symbols, and form the frequency-time UL RE grid. 
\item{} [UL Channel] Evaluate the effect of the channel and form received antenna signal $\mathbf {y}$ at the BS.  The received signal $\mathbf {y}$ is corrupted with additive noise. 
\item{} [UL BS-Rx] Apply a channel estimation algorithm (e.g. LS) to RxData in pilot locations, interpolate over all REs, and store the channel estimate.
\item{} [DL BS-Tx1] Simulate with random parameters: generate transmitted DL bits (possibly including channel coding, such as LDPC), map bits to symbols, and form the frequency-time DL RE grid $\mathbf s$.
\item{} [DL BS-Tx2] Pass the UL-channel estimate to DeepTx. Compute $\mathbf W$ using the output of the DeepTx. 
\item{} [DL BS-Tx3] Calculate $\mathbf x=\mathbf W\mathbf s$ and normalize.
\item{} [DL Channel] Evaluate the effect of the channel and form received antenna signal at the UEs. Again, the signal is corrupted with additive white noise.   
\item{} [UL BS-Rx] For each UE, apply a channel estimation algorithm, equalize the UE RX signal $\mathbf {y}$, and calculate log-likelihood ratios (LLRs) for output bits. 
\end{enumerate}
Steps 1-5 can be computed offline of the main training procedure and saved to a database.

{\bf Remark:} In order to train models that can cope with varying noise levels, it is important that the generated data includes varying levels of noise. Therefore, noise levels in steps [UL Channel] and [DL Channel] should be varied for each sample. In practice, the noise levels for training should be chosen such that the complete range of possible noise levels in the deployment environment is covered (the model might perform poorly if the noise level is outside of its training regime). Since DL SNR is typically somewhat larger than UL SNR due to the higher transmit power of the BS, our approach is to draw the UL SNR (dB) for each sample from a uniform distribution and then add a uniformly distributed random increment to obtain DL SNR for the sample (see \ref{table:param}). 

Training of a neural network is typically carried out by minimizing a loss function. A common choice for binary outputs (bits) is the binary cross entropy (CE) loss \cite{honkala21,korpi21}. For the $q$th training sample, the CE loss can be written as
\begin{align}
  {\mathrm{CE}_q}(\boldsymbol{\theta})=
  - \frac{1}{\#\mathcal DB}\sum_{(i,j)\in \mathcal D}\sum_{l=0}^{B-1} & \left(b_{ijl} \log(\hat{b}_{ijl}) + (1 - b_{ijl}) \log(1 - \hat{b}_{ijl})\right)
\end{align} 
where $\#\mathcal D$ is the number of resource elements carrying data, $B$ is the number of bits, and $\hat{b}_{ijl}=\operatorname{sigmoid}(L_{ijl})$ where $L_{ijl}$ are log-likelihood ratios for decoded bits (in our case, the outputs of the UE receivers). Furthermore, based on our earlier work \cite{honkala21,korpi21}, we have noticed that the results are further improved by weighting the samples based on their SNRs, which gives us the final loss function as
\begin{align}\label{eq:weights}
L_q \left(\bm{\theta}\right) = \operatorname{log}_2\left(1+\mathrm{snr}_q\right) \mathrm{CE}_q \left(\bm{\theta}\right).
\end{align}
where $\mathrm{snr}_q$ is the linear SNR of the $q$th sample.

We, however, found out that the weighted CE loss alone is not sufficient for training DeepTx. For example, training a model for varying prediction lengths ($\tau$) is difficult and training a model with random initialization leads to suboptimal performance. In such a case, a model with a sufficient performance can be learned by initializing the weights using a model trained for a single gap ($\tau_{\max}=1$). Considering a multi-user (MU) scenario introduces further difficulties as it was determined that a model with sufficient performance was only found by starting from a pre-trained single user (SU)-model. This means that a three-stage training is required to obtain a sufficient multi-user model which can handle a varying prediction length (i.e., SU / $\tau_{\max}=1$ model $\rightarrow$ MU / $\tau_{\max}=1$ model $\rightarrow$ MU /  $\tau_{\max}>1$ model).

To avoid this difficult and time-consuming multistage training procedure, we found out that the exponential (EXP) loss, 
\begin{equation}
    {\mathrm{EXP}}(\boldsymbol{\theta})= \frac{1}{\#\mathcal DB}\sum_{(i,j)\in \mathcal D}\sum_{l=0}^{B-1}\exp(-y  \hat y) , \quad y = 2b_{ijl} - 1,   \hat y = 2 \hat{b}_{ijl} - 1,
  \end{equation}
  provides a promising alternative for CE. In our experiments, the EXP loss allowed us to train models from random initialization (i.e. without pre-training) with good performance for both SU and MU scenarios.
Our speculation is that the EXP loss is helping optimization here as it penalizes heavily wrongly classified samples and especially outliers, thereby pushing the training faster to right regime. 
The EXP loss, however, has an disadvantage: as it is not based on any probabilistic derivations, there is no warranty that the model output would approximate log-likelihood ratios (needed, e.g., for decoding). To overcome this issue, we trained models starting with the EXP loss and then switching to the CE loss with a linear transition period at the end of the training. 
Furthermore, the loss of each sample (for both EXP or CE losses) is weighted based on its (downlink) SNR as in \eqref{eq:weights}.

\subsection{Approximated ZF-transformation}
\label{sec:appr-zf-transf}

The ZF transformation calculated using Eq. \eqref{eq:ZF} involves a matrix inversion which can be computationally inefficient.
A computationally more efficient alternative is to approximate the inverse with a series expansion \cite{wu2013 ,zhu15}, 
\begin{displaymath}
\mathbf A^{-1}=(\mathbf D+ \mathbf E)^{-1}=\sum_{n=0}^\infty(-\mathbf D^{-1} \mathbf E)^n \mathbf D^{-1} 
\end{displaymath}
where $\mathbf D$ is the diagonal part of $\mathbf A$ and $\mathbf E=\mathbf A-\mathbf D$. If we choose $\mathbf A=\mathbf H_{ij}\mathbf H_{ij}^*$ or
\begin{displaymath}
  \mathbf D= \mathrm{diag}(\mathbf H_{ij}\mathbf H_{ij}^*)\quad\mathrm{and}\quad \mathbf E = \mathbf H_{ij}\mathbf H_{ij}^* - \mathrm{diag}(\mathbf H_{ij}\mathbf H_{ij}^*),
\end{displaymath}
we can write an approximation 
\begin{displaymath}
  \mathbf W_{ij}  \approx \mathbf H_{ij}^*\sum_{n=0}^{k-1}(-\mathbf D^{-1} \mathbf E)^n \mathbf D^{-1} . 
\end{displaymath}
The efficiency can be further improved by computing the product $\mathbf W_{ij}\mathbf s_{ij}$ directly without forming $\mathbf W_{ij}$:
\begin{equation}\label{eq:ZFapprox}
  \mathbf x_{ij}= \mathbf W_{ij}\mathbf s_{ij} \approx
  \mathbf H_{ij}^* \left[\mathbf z_{ij} + (-\mathbf D^{-1} \mathbf E) \mathbf z_{ij} +\ldots+ (-\mathbf D^{-1} \mathbf E)^{k-1} \mathbf z_{ij}\right], 
\end{equation}
where $\mathbf z_{ij}=\mathbf D^{-1} \mathbf s_{ij}$. The computation of the above series approximation essentially requires repeated multiplication of a vector with $\mathbf E$ and then dividing it with the diagonal elements of $\mathbf D$.

\section{Simulation results}
\label{sec:validation_res}

We evaluate our methods using simulated 5G 4x2 MIMO experiments considering both SU and MU scenarios. The simulation parameters are listed in Table \ref{table:param}. We mostly consider a DMRS configuration which contains pilots in the 3rd and 12th OFDM symbols, with even (odd) subcarriers allocated for the pilots of first (second) layer. However, as beamforming is often performed using SRS pilot signals transmitted in UL, we also show results where the channel is estimated using SRS type reference signals consisting of pilot symbols in the second to last OFDM symbol (DL still utilizes the 2-pilot DMRS for data demodulation at the UEs).

Channel realizations for training and validation data are generated with a link-level simulator implemented with Matlab's 5G Toolbox \cite{Matlab5G}. Parameters for each channel realization are chosen randomly and each channel realization includes  10 TTIs. The database contains 22374 channel realizations for training and 10612 for validation. As described in Section \ref{sec:training_data}, at least DL needs to be simulated during training procedure. We simulate both UL and DL during training using a Tensorflow implementation of the system model described in Section \ref{sec:simulator}, with channel realizations randomly picked from the generated channel databases. Note that the simulations do not include any hardware impairments as the focus of this work is in evaluating the performance of DeepTx in dealing with the propagation channel effects. Introducing hardware non-idealities into the simulations is an important future work topic for us, as this will result in the effective channel becoming slightly non-reciprocal, providing more opportunities for ML-based beamforming solutions to improve upon conventional systems.

The model architecture is described in Section \ref{sec:nn_tx} (Table~\ref{tab:cnn_arch}). The training of DeepTx is carried out by minimizing first the EXP loss and then switching to CE loss as described in Section \ref{sec:training_data} (the linear transition period is between 90\% and 95\% of iterations). As in \cite{honkala21}, we employ the LAMB optimizer \cite{You2020Large} using a learning rate scheduler that has first a linear warm up period (1600 iterations) and ending with the linear decay period (starting after 30\% of iterations). The models are trained using multi GPU systems with total batch size of 72. The learning rate is set to $3.6\cdot 10^{-5}$ and the number of samples iterated (with replacement) during training is $35\cdot 10^{6}$. These parameters were found experimentally.

We did not observe overfitting with our models in most of the cases. The only exception is the 1 pilot configuration with no additional UL history data, in which case the input to DeepTx only contains information about channel over one symbol, giving no insight about possible evolution of the channel.  In this case we observed some level of overfitting (reduction of performance in validation statistics) especially with large models, which might be due to insufficient information in the input, resulting in the DeepTx model to start memorizing training data.

The results are compared to two baselines: ZF precoding with unprocessed UL channel estimate (Section \ref{sec:trad-rece-proc}) and ZF using full channel information about DL channel in question. The former represents practical baseline while the latter approaches the theoretical upper bound of performance. The metric for performance is DL bit error rate (BER) achieved by the LMMSE-based UE receiver(s). Note that channel coding is omitted from this analysis for simplicity, as we do not expect it to impact the relative performance of the different schemes.

\begin{table}
  \setlength{\tabcolsep}{3pt}
    \renewcommand{\arraystretch}{1.3}
    \footnotesize
    \centering
    \caption{Simulation parameters for the numerical simulation.}
    \begin{tabular}{l|p{5.4cm}|l}
    \hline
    \multicolumn{1}{c|}{\textbf{Parameter}} & Value & \textbf{Randomization}\\
		\hline\hline
		Carrier frequency & 4 GHz & None\\
    \hline
		Channel model & TDL-A, TDL-B, TDL-C  & Uniform\\
		\hline
		RMS delay spread & 10 ns -- 300 ns & Uniform\\
		\hline
		UE velocity range & 0 -- 30 km/h & Uniform\\  
		\hline
               SNR UL & $0$ dB -- $35$ dB & Uniform\\
		\hline
               SNR DL & SNR UL + $\Delta$, $\Delta=1$ dB -- $5$ dB& Uniform\\
                \hline
		Number of PRBs & 14 (168 subcarriers) & None\\
		\hline
		Subcarrier spacing & 30 kHz & None\\
		\hline
		TTI length & 14 OFDM symbols / 0.5 ms & None\\
		\hline
		Modulation scheme & 16-QAM & None\\
		\hline
    Number of MIMO layers&2&None\\
    \hline
		Number of BS antennas & 4 & None\\
		\hline
		Number of UE antennas & 2 per UE (SU scenario) or 1 per UE (MU scenario)  & None\\
		\hline
    \end{tabular}
    \label{table:param}
  \end{table}

\subsection*{Primary validation results}

\begin{figure*}%
	\centering
        \hspace{-0.4cm}
	\subfloat[SU, $\tau=1$]{{\hspace{-0.4cm}\includegraphics[width=0.54\columnwidth]{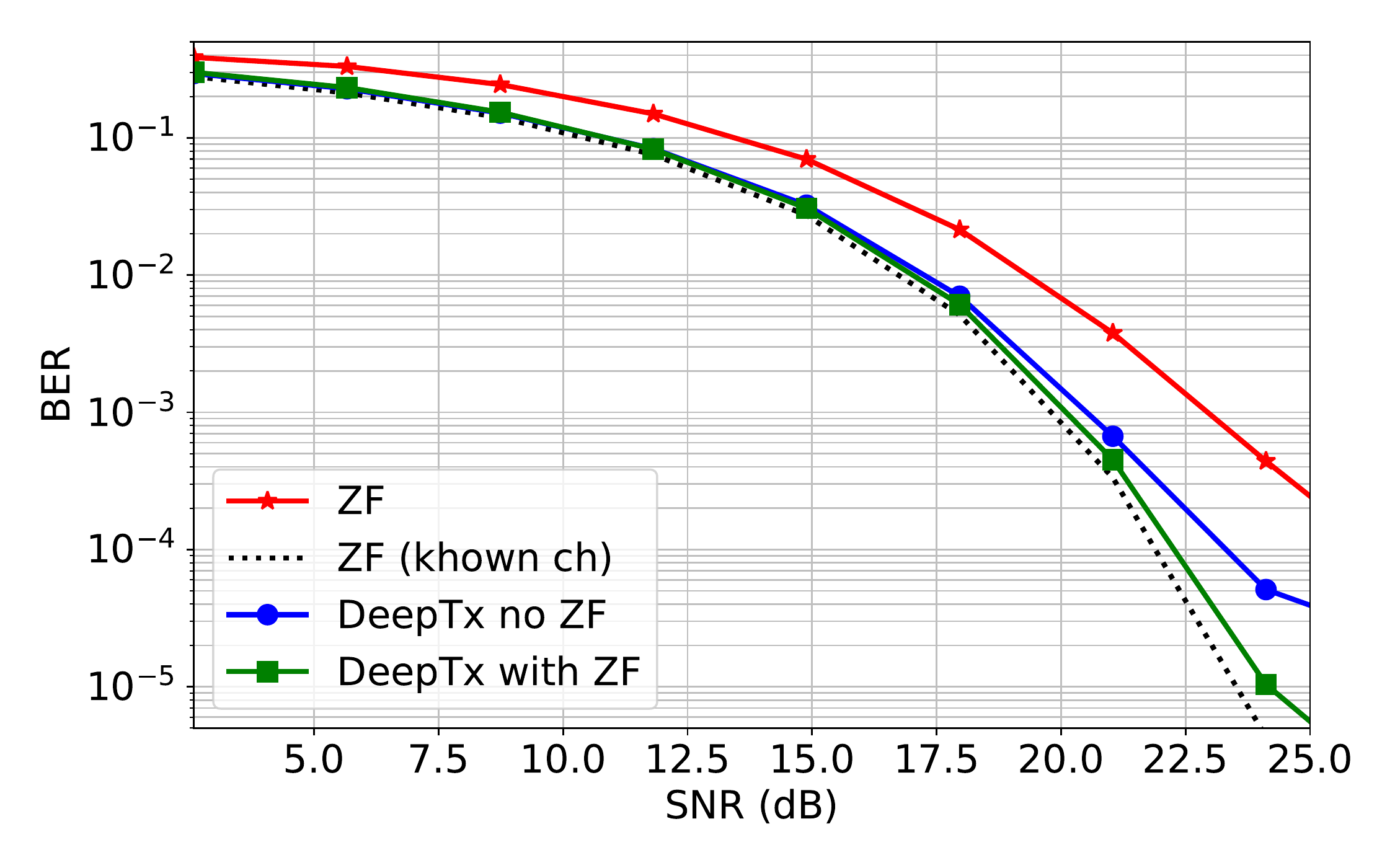} }}%
	\subfloat[MU, $\tau=1$]{{\hspace{-0.3cm}\includegraphics[width=0.54\columnwidth]{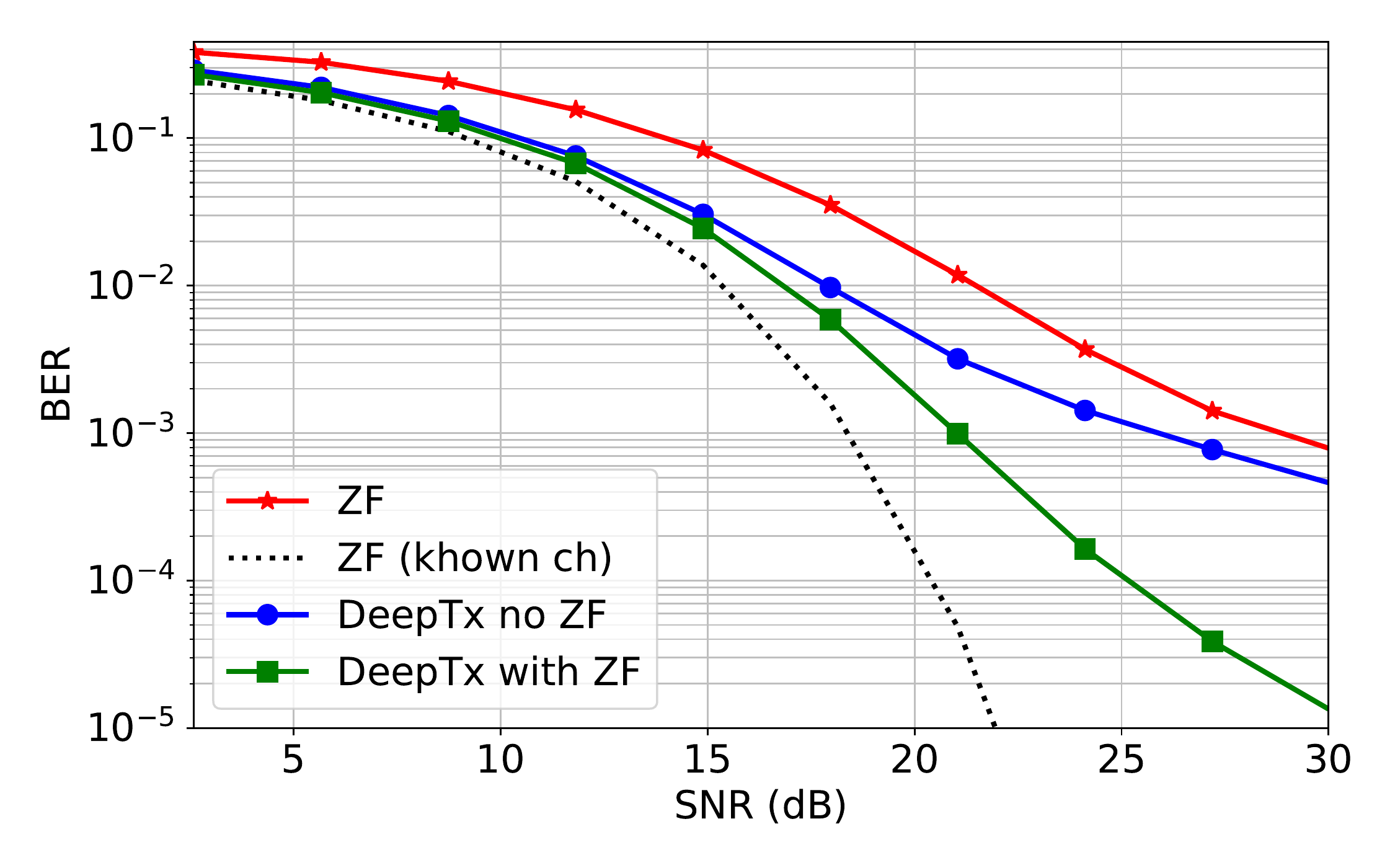} }}%
        \\\vspace{-0.4cm}
	\subfloat[SU, $\tau=3$]{{\hspace{-0.4cm}\includegraphics[width=0.54\columnwidth]{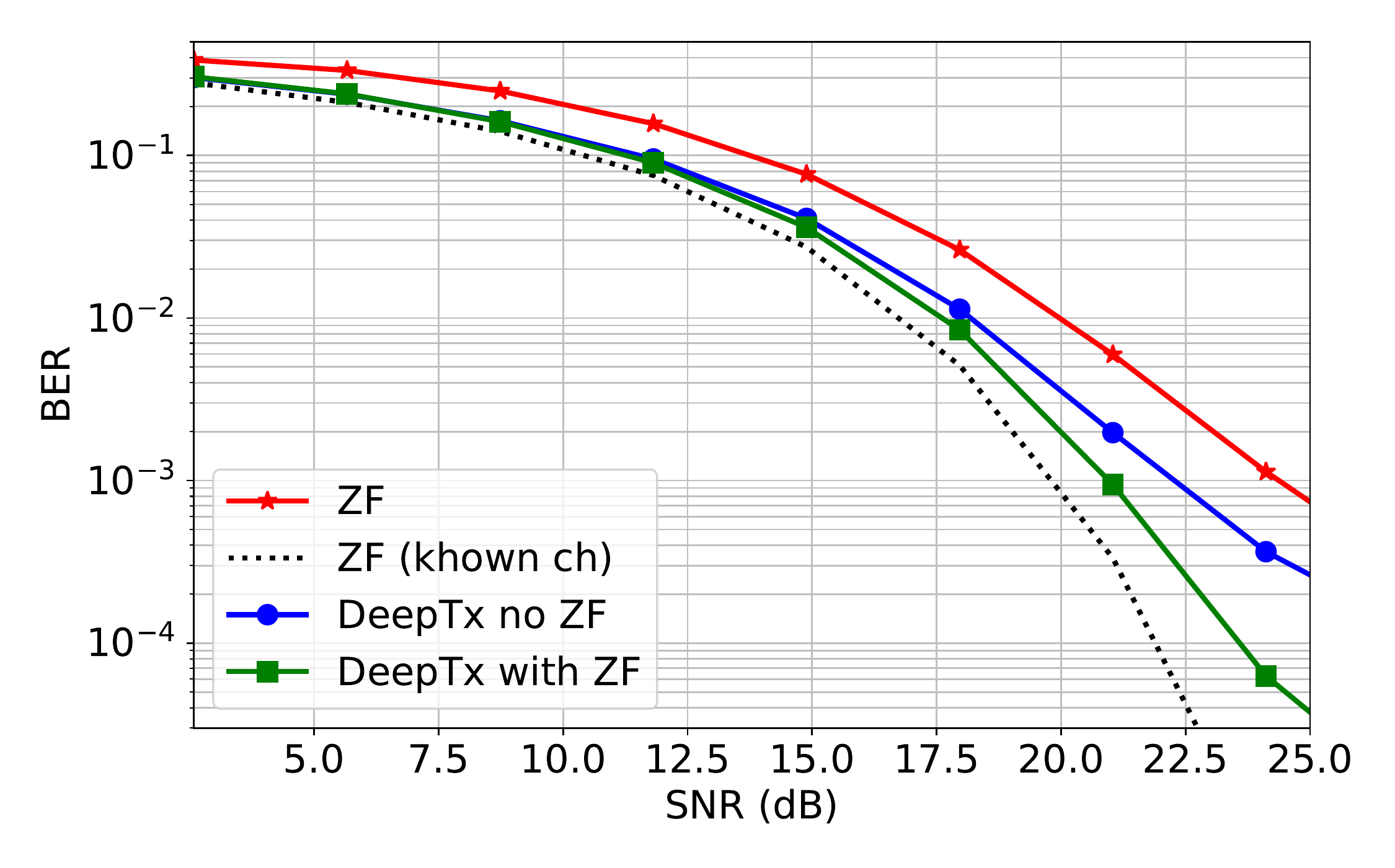} }}%
	\subfloat[MU, $\tau=3$]{{\hspace{-0.3cm}\includegraphics[width=0.54\columnwidth]{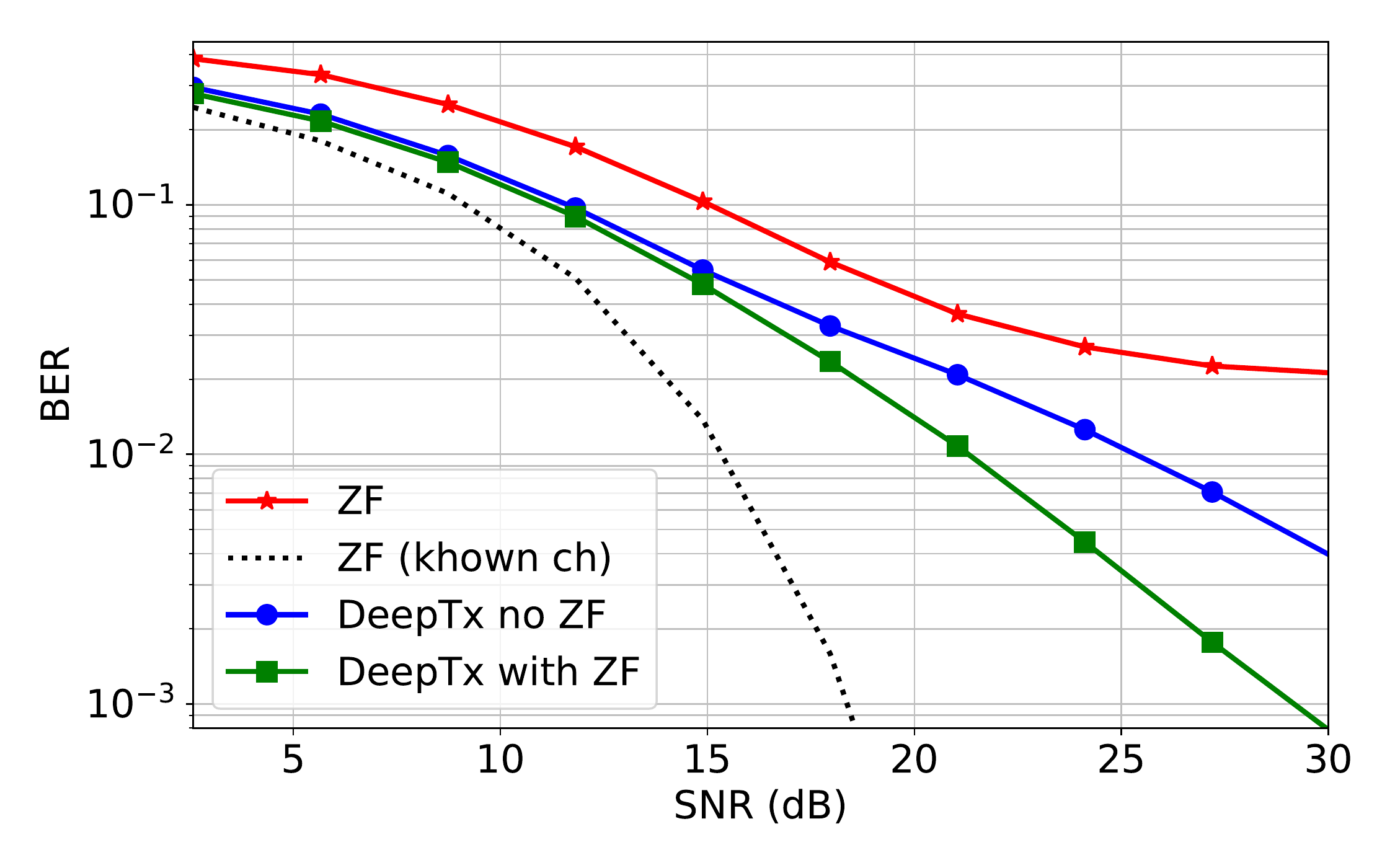} }}%
	\\\vspace{-0.4cm}
        \subfloat[SU, $\tau=6$]{{\hspace{-0.4cm}\includegraphics[width=0.54\columnwidth]{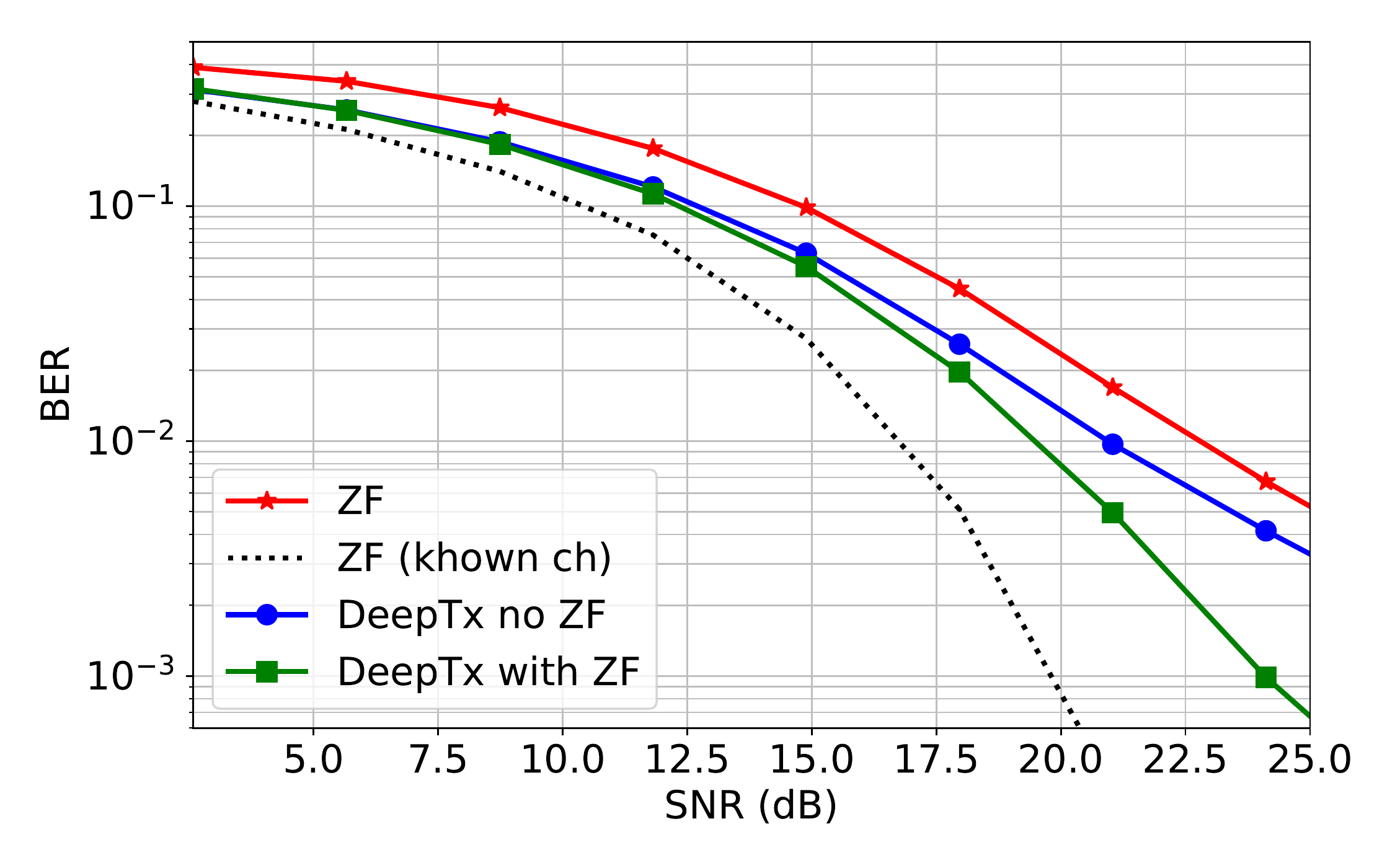} }}%
	\subfloat[MU, $\tau=6$]{{\hspace{-0.3cm}\includegraphics[width=0.54\columnwidth]{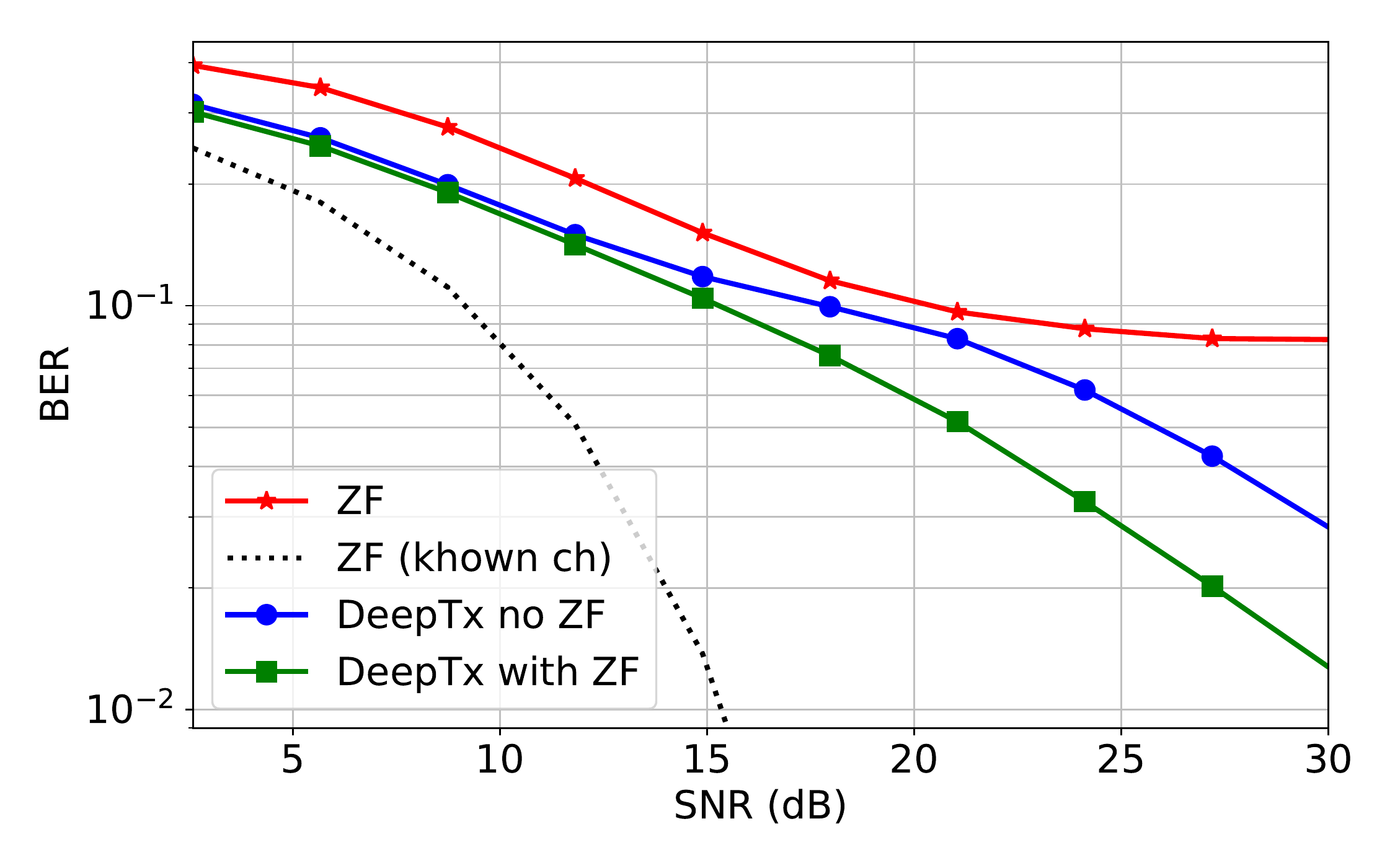} }}%
	\caption{Uncoded BER for LMMSE receiver in UEs when DeepTx+ZF (blue) or ZF (red) as precoding, ranging for prediction lengths $\tau = 1, 3, 6$; single-user scenario (a), (c), (e); multi-user scenario  (b), (d), (f). The results are for 2 pilot DMRS configuration.}%
	\label{fig:deeptx_results}%
\end{figure*}

First, we study performance of DeepTx both in SU and MU scenarios. First, we consider the 2-pilot DMRS configuration as such a pilot configuration can provide information about the evolution of the channel within a single TTI. Figure~\ref{fig:deeptx_results} shows results for DeepTx both with and without explicit ZF transformation. It can be seen that DeepTx clearly outperforms the ZF baseline in both SU and MU scenarios. In addition, combining DeepTx with the ZF transformation gives also a significant improvement compared to DeepTx which has to learn the proper beamforming transformation from scratch. This can be expected, as without explicit ZF, significant part of ResNet capacity is likely used for calculating the precoding matrix, which is an inherently nonlinear task, compared to DeepTx with explicit ZF in which case the ResNet only needs to carry out prediction of channel evolution.

Prediction of the channel is especially difficult when there are pilots only in one OFDM symbol per TTI (e.g., SRS), in which case the LS channel estimate cannot provide any knowledge about the evolution of the channel. The results for the SRS-based scenario in MU case are presented in Figs.~\ref{fig:history_results}~(a), (c), and (e). The performance of DeepTx is still better than the baseline under low SNRs, but it reaches only similar performance with high SNRs. One reason for the higher performance in lower SNRs could be that DeepTx acts as a pre-processing step for ZF to provide improved beamforming, for example, through better noise tolerance.

\subsection*{Effect of adding history of channel estimates}

\begin{figure*}%
	\centering
	\subfloat[2 pilot, $\tau=1$]{{\hspace{-0.4cm}\includegraphics[width=0.54\columnwidth]{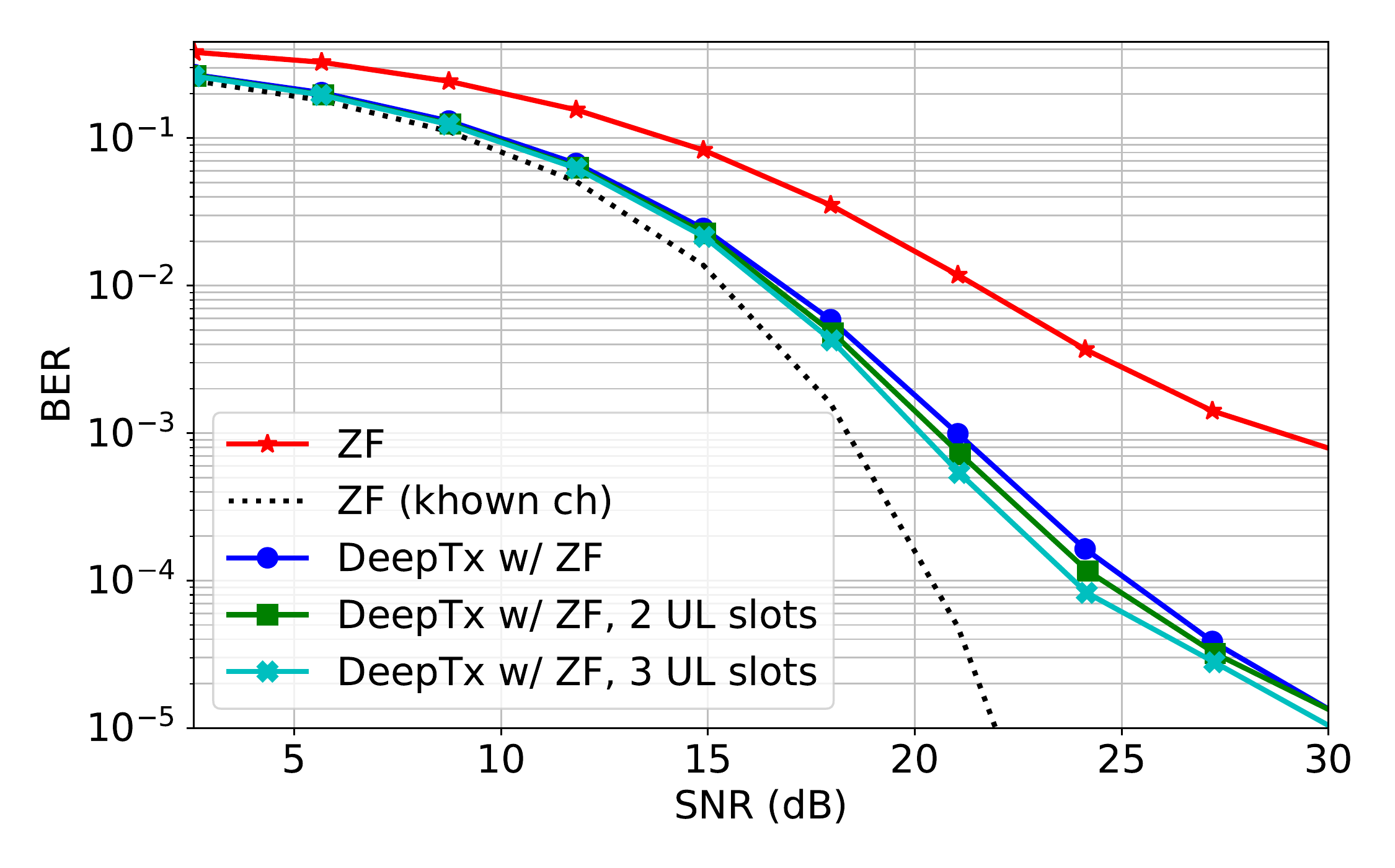} }}%
	\subfloat[SRS pilot, $\tau=1$]{{\hspace{-0.3cm}\includegraphics[width=0.54\columnwidth]{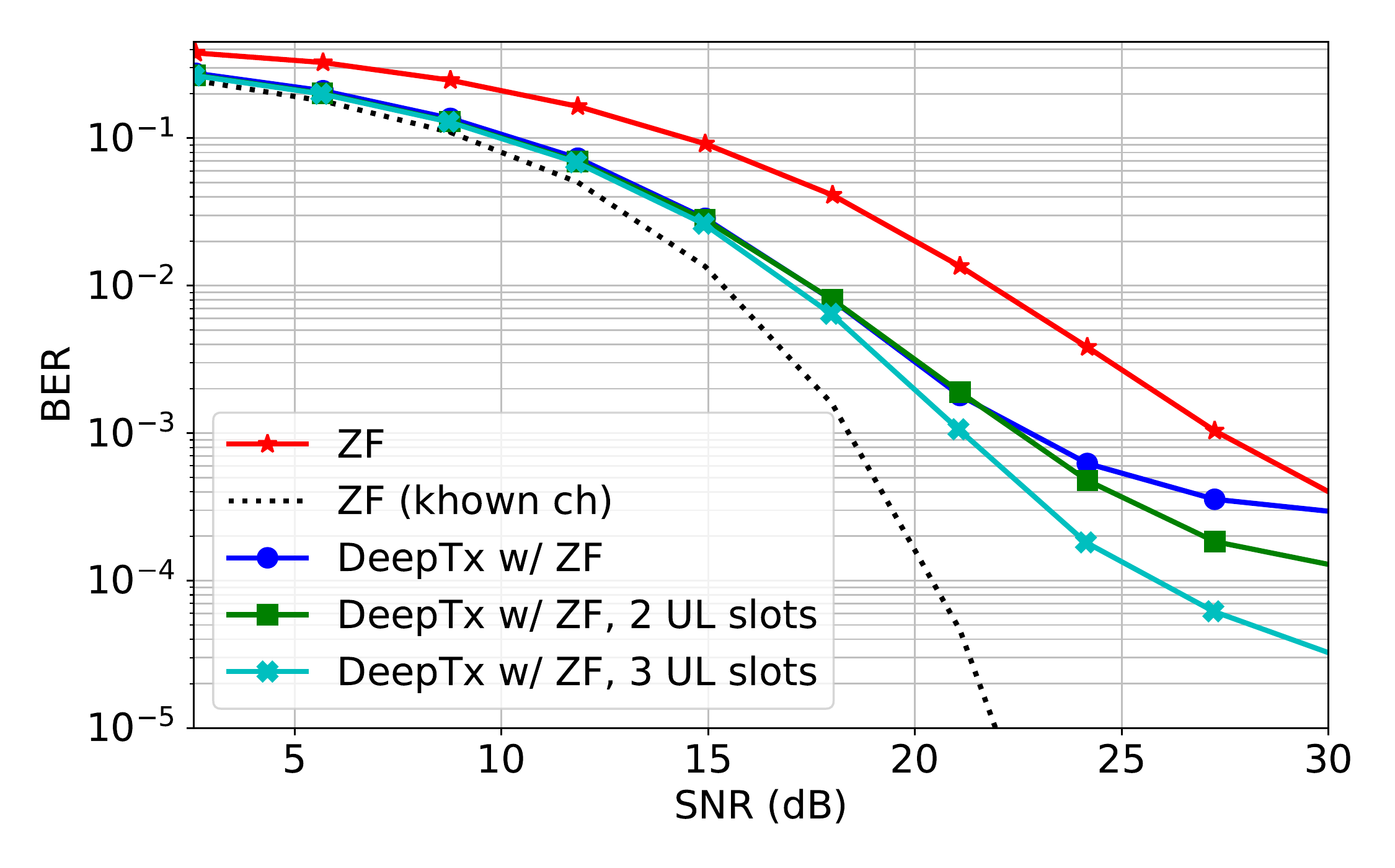} }}%
        \\\vspace{-0.4cm}
	\subfloat[2 pilot, $\tau=3$]{{\hspace{-0.4cm}\includegraphics[width=0.54\columnwidth]{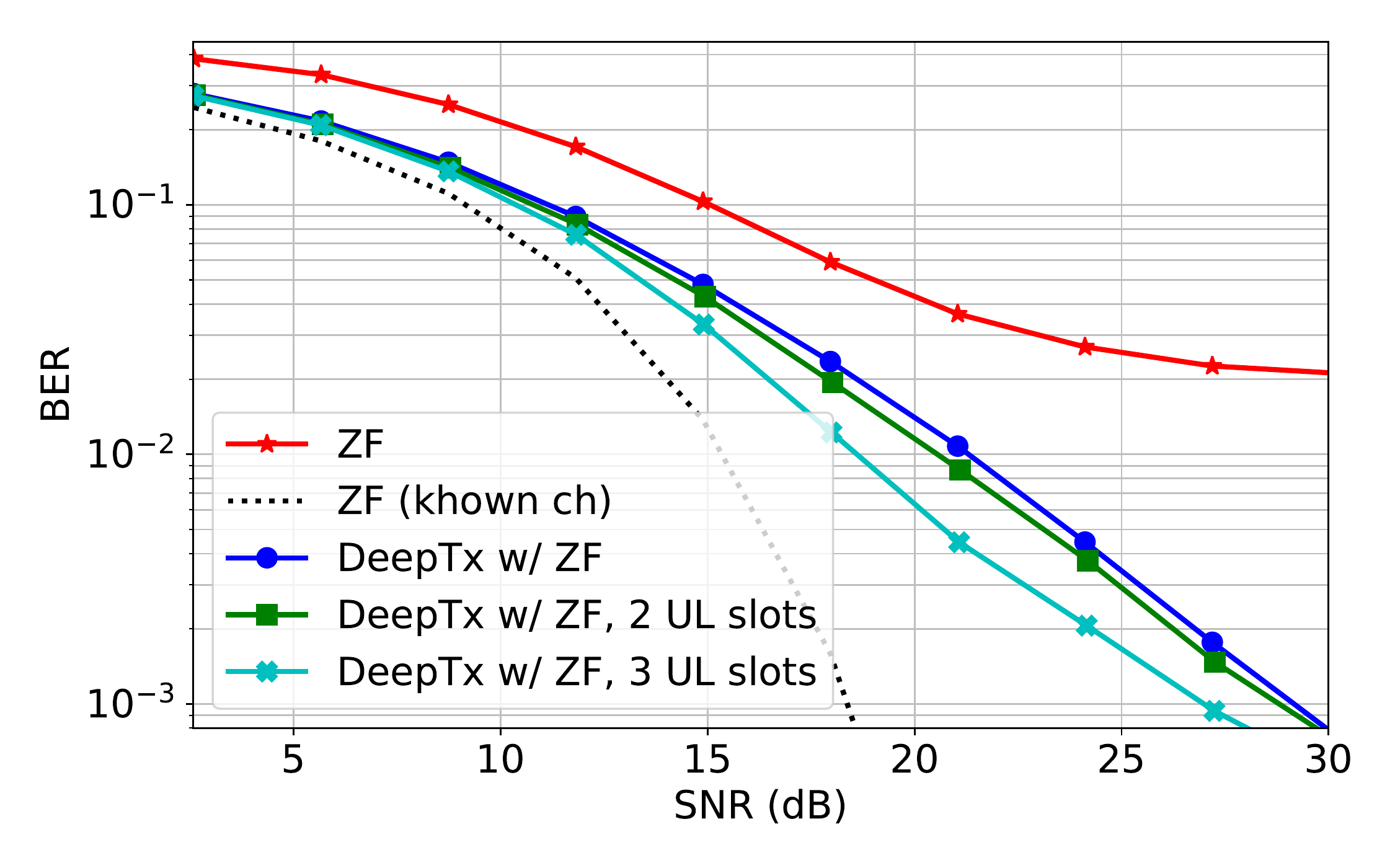} }}%
	\subfloat[SRS pilot, $\tau=3$]{{\hspace{-0.3cm}\includegraphics[width=0.54\columnwidth]{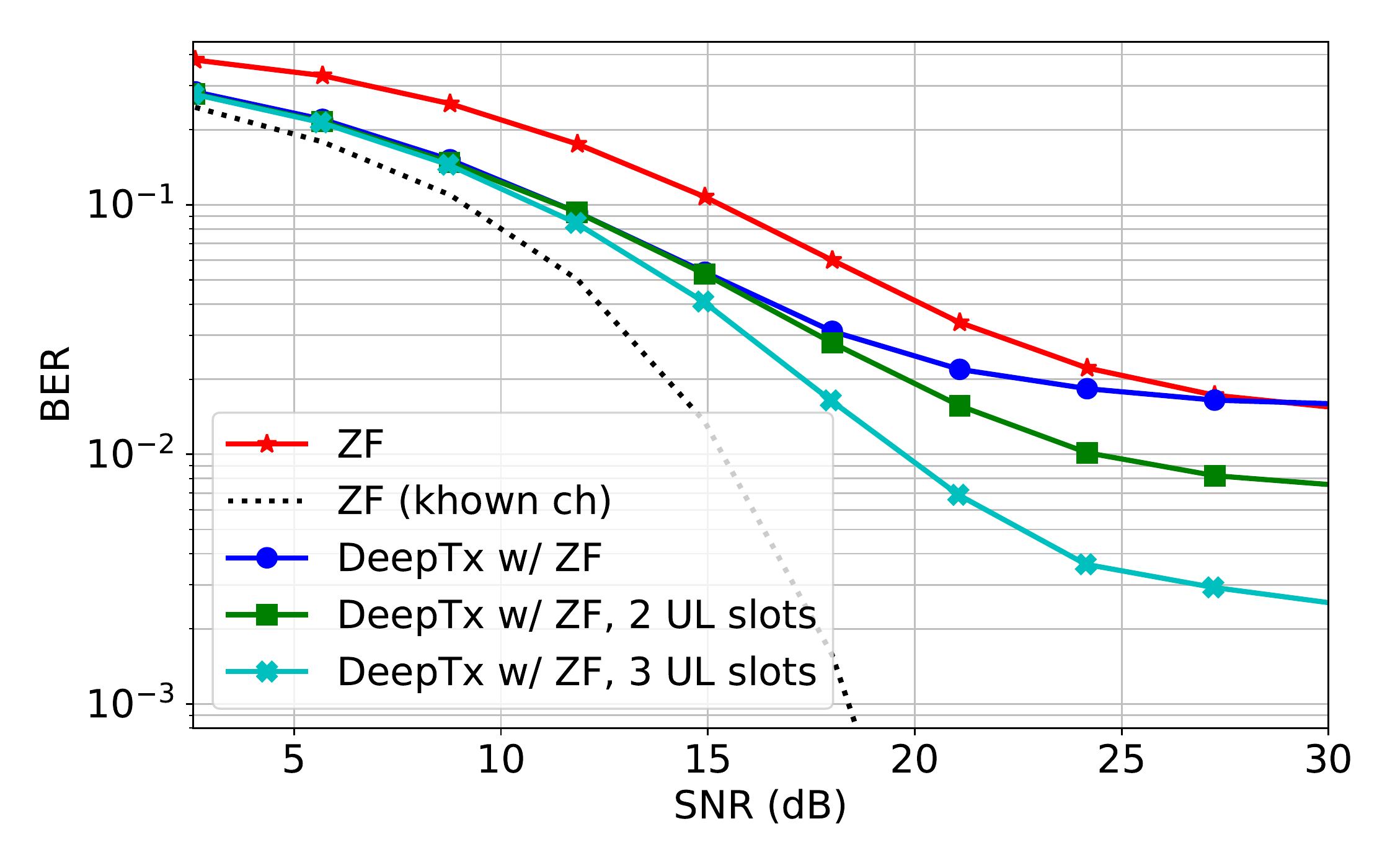} }}%
	\\\vspace{-0.4cm}
	\subfloat[2 pilot, $\tau=6$]{{\hspace{-0.4cm}\includegraphics[width=0.54\columnwidth]{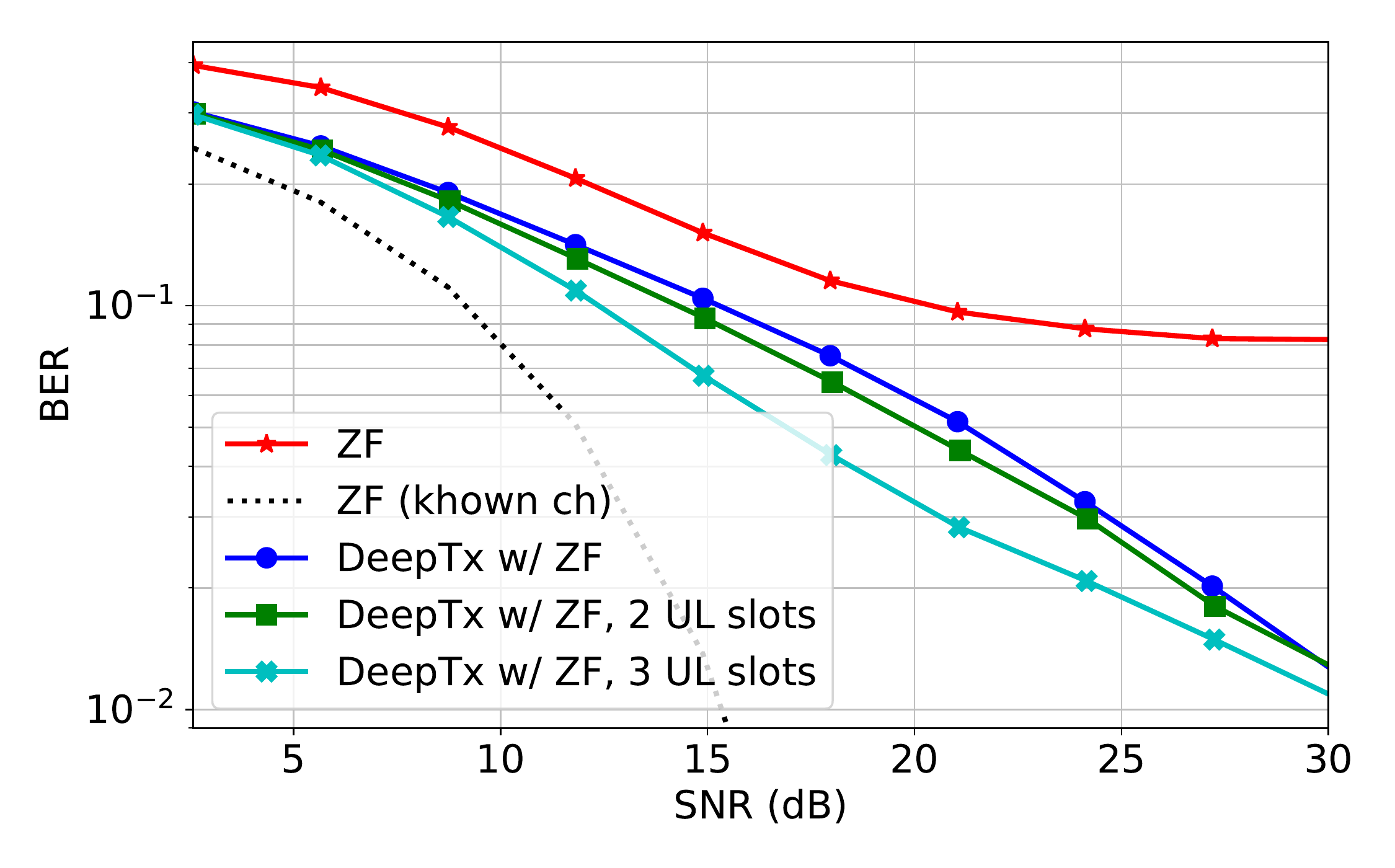} }}%
	\subfloat[SRS pilot, $\tau=6$]{{\hspace{-0.3cm}\includegraphics[width=0.54\columnwidth]{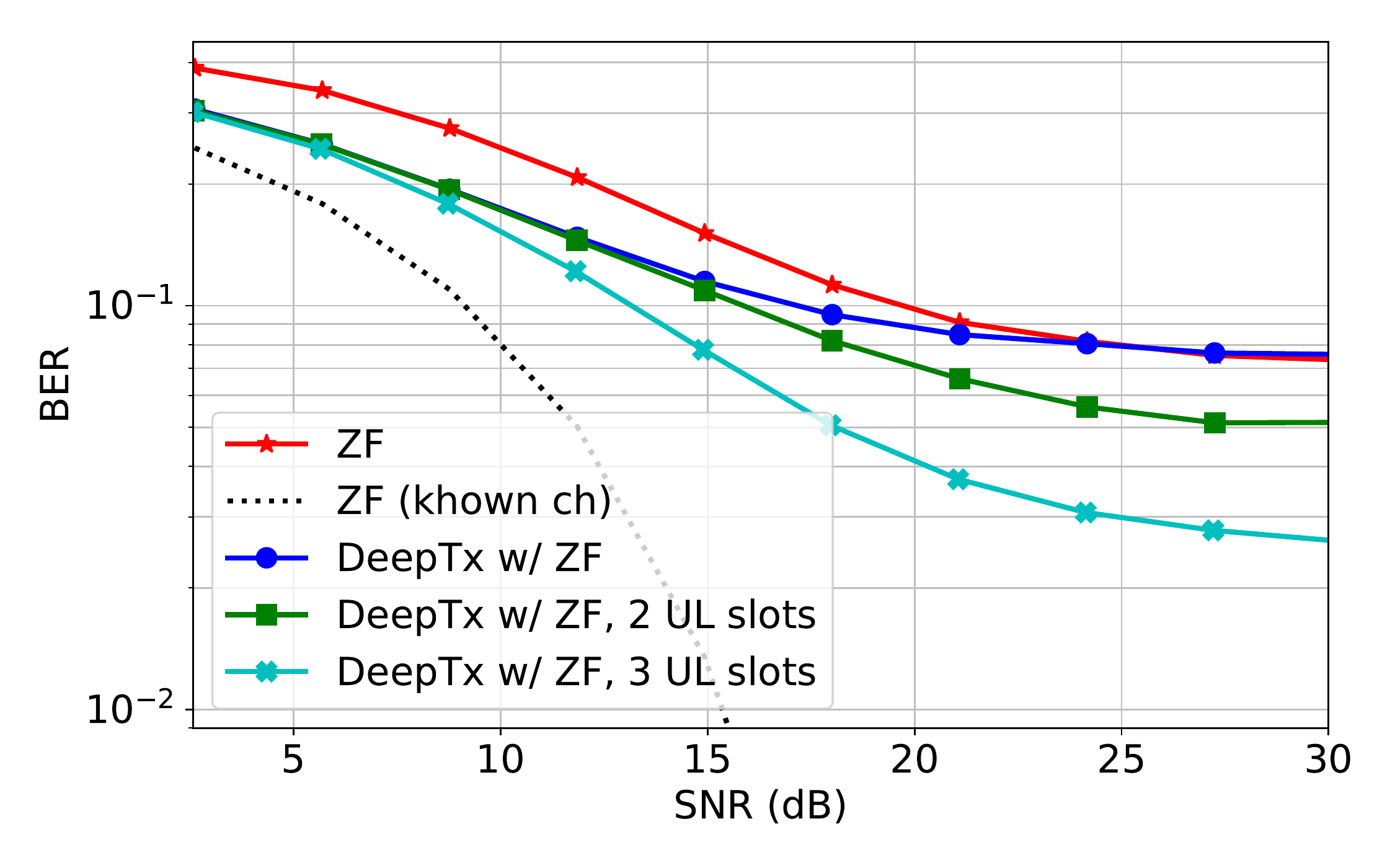} }}%
	\caption{DeepTx results for MU scenario, including models which get more then one UL slot as input. Results are shown for both 2-pilot DMRS and SRS type UL pilot configurations. \label{fig:history_results}}
\end{figure*}

It is also possible to utilize channel information from history. To this end, we consider a case in which channel estimates are obtained for up to three adjacent UL slots before DL slots.\footnote{This might not be a practical setup as UL slots are typically more spread out in time, but serves as an initial experiment. We will consider cases in which gaps between UL slots are larger and more sporadic in future research.} The architecture for multi-UL-slot DeepTx is similar to Table \ref{table:param}, except that the channel estimates for different slots are stacked (the last dimension) and then provided as input to DeepTx. Training procedure is also similar to that described in Section \ref{sec:training_data}, but UL processing (steps 2-4) are repeated for all UL slots. Separate models are trained for three cases, where input is either 1, 2, or 3 UL slots. The results are shown in Figure \ref{fig:history_results}. It can be seen that performance for the SRS-based scenario is significantly improved when more UL slots are given as input. This can be expected as more information about the evolution of the channel is delivered to the model. A significant improvement can also be observed in the 2 pilot case, especially with larger values of $\tau$.

\subsection*{Approximation of ZF transformation}

\begin{figure*}
	\centering
	\subfloat[$\tau=1$]{{\hspace{-0.4cm}\includegraphics[width=0.54\columnwidth]{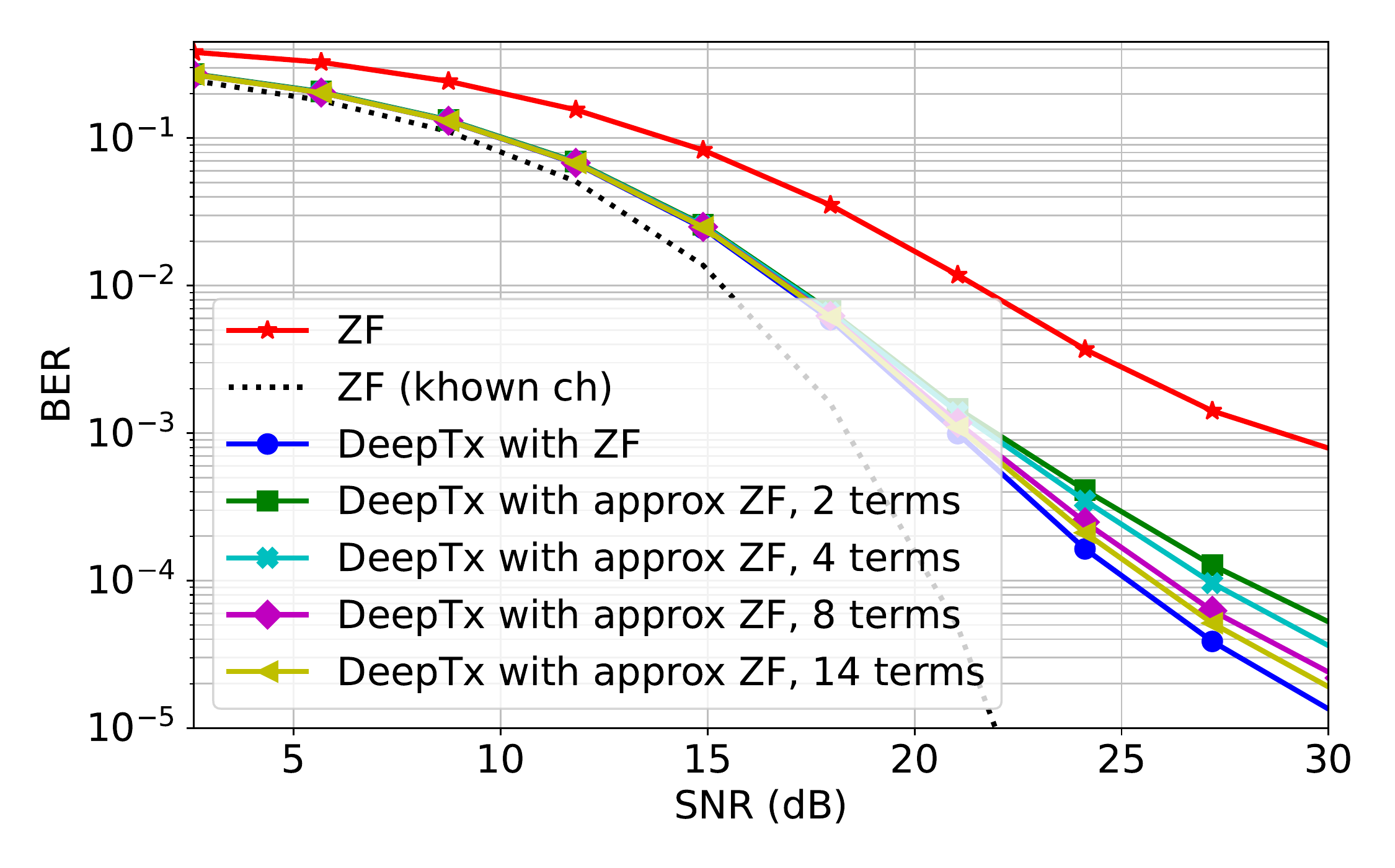} }}%
	\subfloat[$\tau=6$]{{\hspace{-0.4cm}\includegraphics[width=0.54\columnwidth]{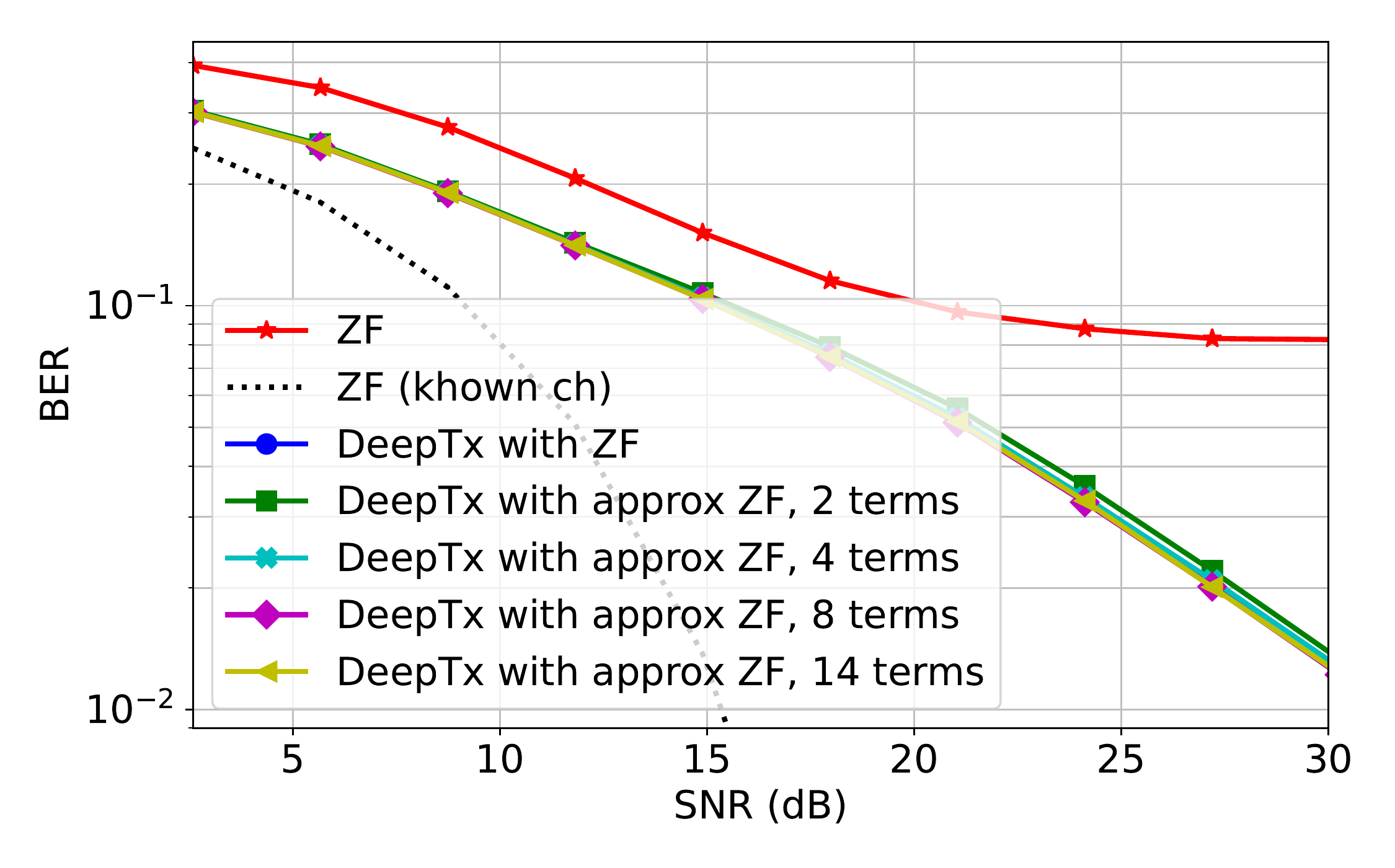} }}%
         \vspace{-0.4cm}
	\caption{The results for DeepTx with the approximated ZF, (a) for $\tau = 1$, and (b) for  $\tau = 6$.}
	\label{fig:zf_approx}%
\end{figure*}

As described in Section \ref{sec:appr-zf-transf}, the computational complexity of ZF transformation can be reduced by replacing the matrix inversion with a series approximation. The performance of DeepTx with approximated ZF with different number of terms is reported in Figure \ref{fig:zf_approx}. For longer prediction periods, four terms provides almost no performance loss compared to the exact ZF transformation and the performance with two terms (i.e., $\mathbf W_{ij} \approx \mathbf H_{ij}^* (I -\mathbf D^{-1} \mathbf E) \mathbf D^{-1}$) is only slightly degraded. There is a minor performance loss for short prediction periods (e.g. $\tau=1$) under higher SNRs. We also noticed experimentally that odd number of terms in the approximation leads to insufficient performance. For example, the approximation with 21 terms is significantly worse than performance with just two terms.

\subsection*{Benefits of training DeepTx based on beamforming performance}

\begin{figure*}%
	\centering
	\includegraphics[width=0.73\columnwidth]{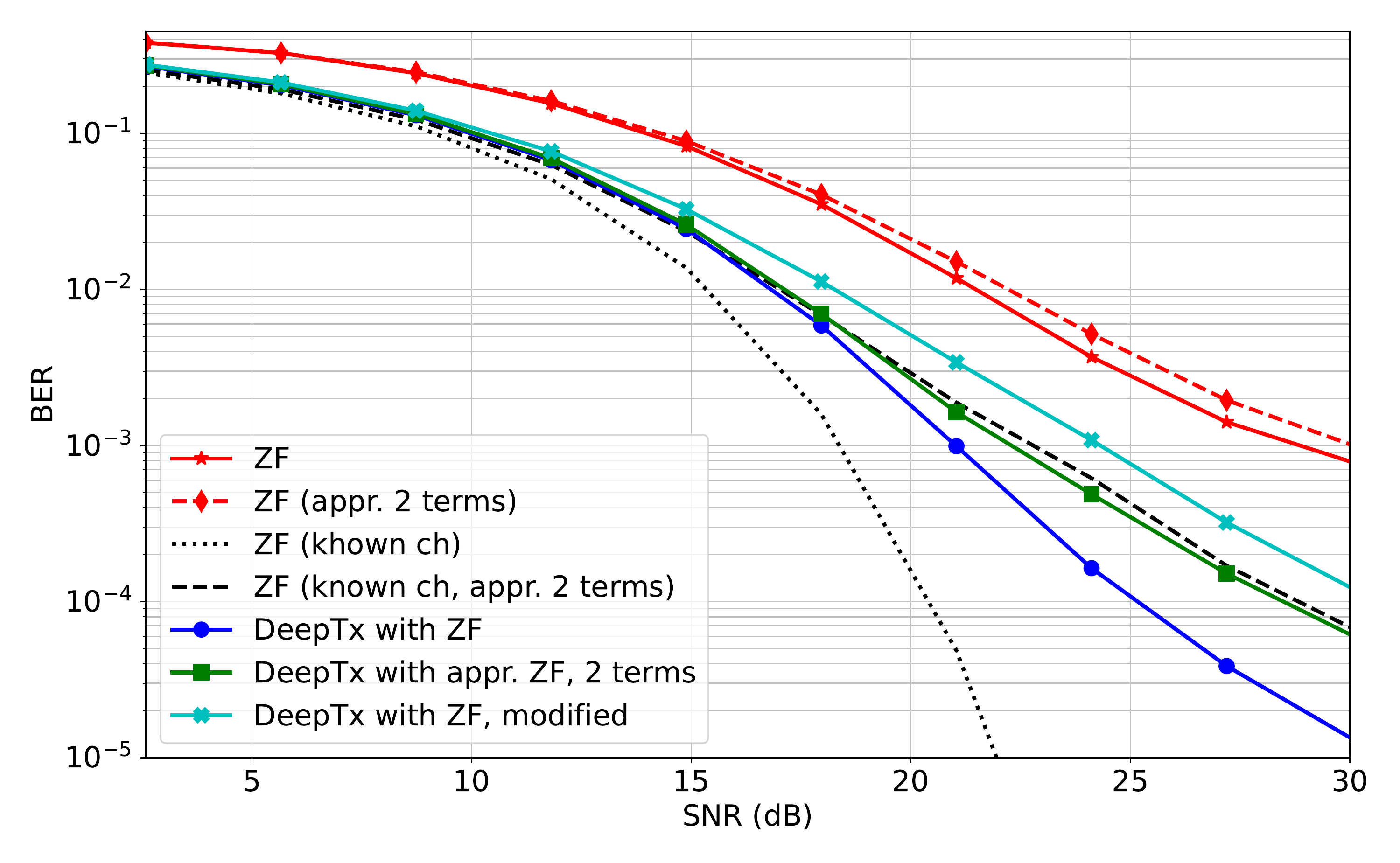}
	\caption{Results with approximated ZF transformations. The modified DeepTx model (cyan) is the model which is trained using the exact ZF transformation, but ZF is replaced by its two term approximation in the testing phase.  The figure also shows baselines in which the approximated ZF is used instead of the exact ZF. The results are for $\tau=1$.}
	\label{fig:zf_mod_approx}%
      \end{figure*}

One of the potential benefits of the proposed solution is that DeepTx could learn to pre-augment the input to the beamformer to handle errors in the beamforming phase. To study this hypothesis, we carry out an experiment with the DeepTx model that is trained with the exact ZF beamformer, which is then replaced with the two term ZF approximation in the testing phase. This model is compared to the model which is also trained with the two term approximation. Results are shown in Figure \ref{fig:zf_mod_approx}. We can see that training with the approximation leads to significantly higher performance, indicating that DeepTx has learned to handle errors stemming from the approximation.

      We also experimented with a setup in which a DeepTx type network (i.e., a ResNet architecture described in Table~\ref{tab:cnn_arch}) is trained to predict the future DL channel without considering the beamforming. The setup is otherwise similar to the one described above, but the cost function was chosen to be the $L^2$-loss between the network output and the true DL channel. The resulting model provides a better prediction of the DL channel than using the channel estimate from UL (MSE is approximately 3.6 dB lower).  However, when the channel predicted by the model is used for ZF beamforming, the beamforming performance is worse than simply using the UE channel estimate without any prediction. This indicates that the task of producing a suitable input to ZF transformation and training using loss from DL UE performance is better than the prediction of the actual channel (or the DeepTx architecture may not be as well-suited for pure channel prediction).

\subsection*{Experiments with different model architectures and notes about model complexity}

\begin{figure*}%
	\centering
	\subfloat[$\tau=1$]{{\includegraphics[width=0.51\columnwidth]{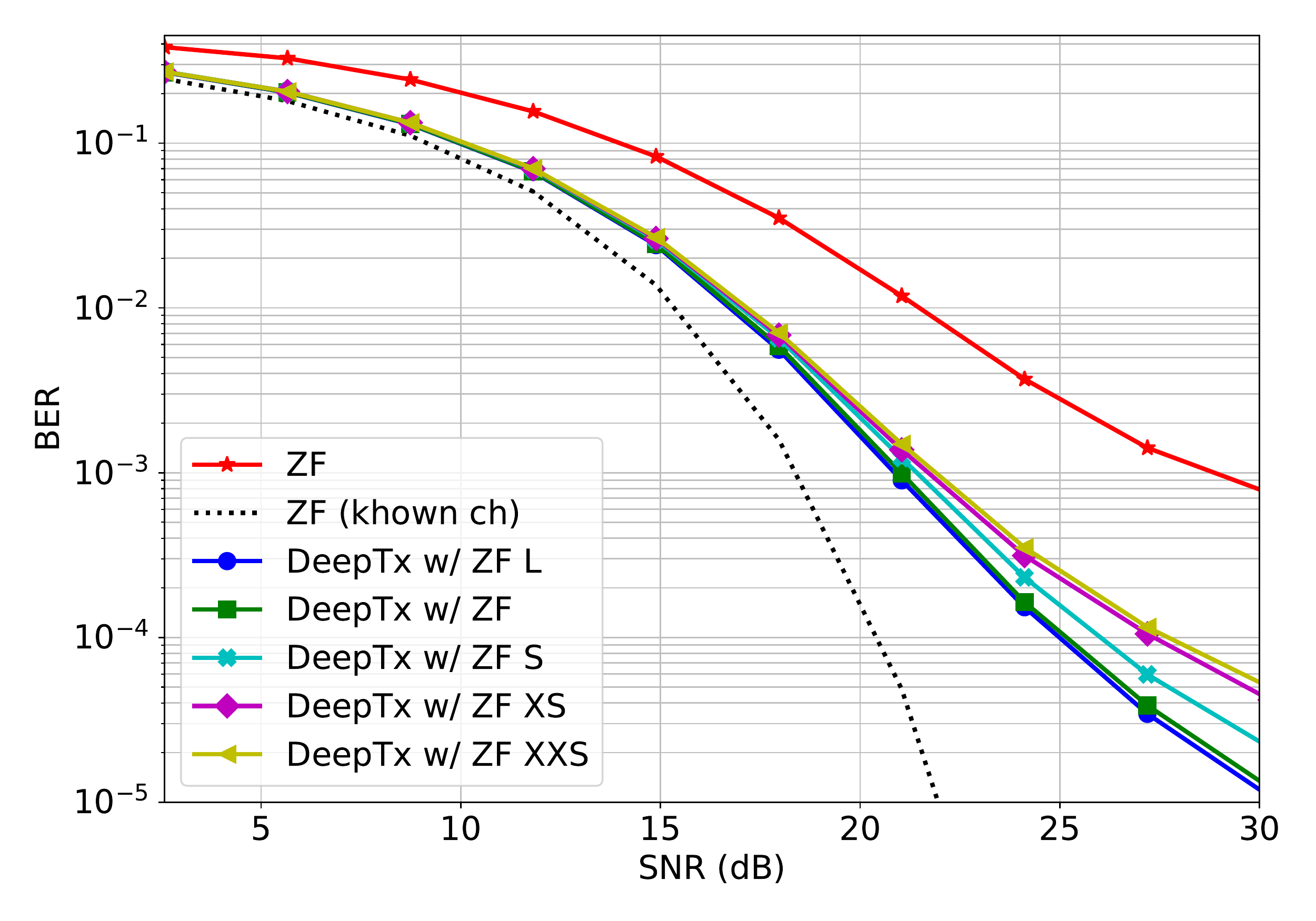} }}%
	~
	\subfloat[$\tau=6$]{{\includegraphics[width=0.51\columnwidth]{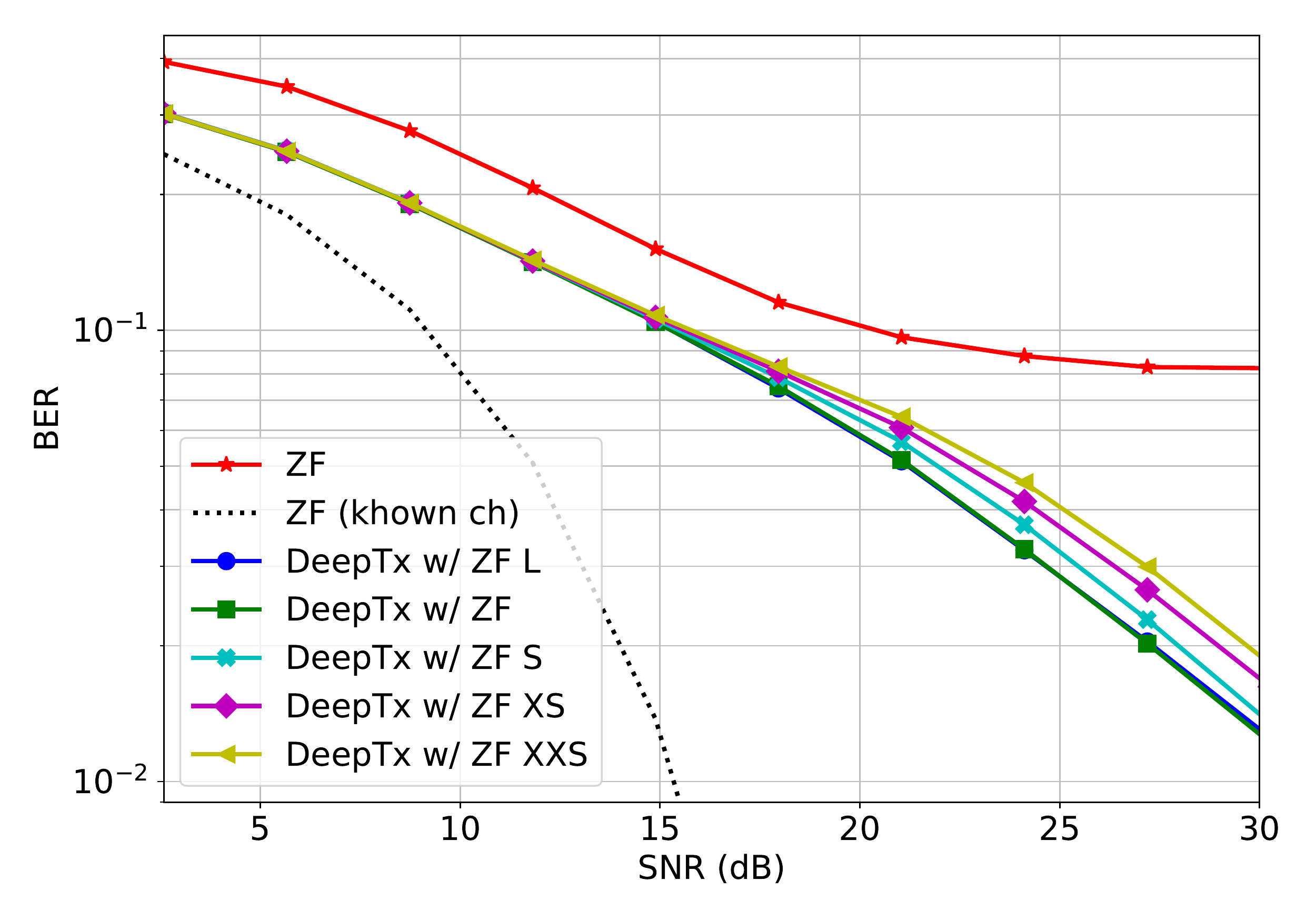} }}%
	\\\vspace{-0.2cm}
	\subfloat[$\tau=1$]{{\includegraphics[width=0.51\columnwidth]{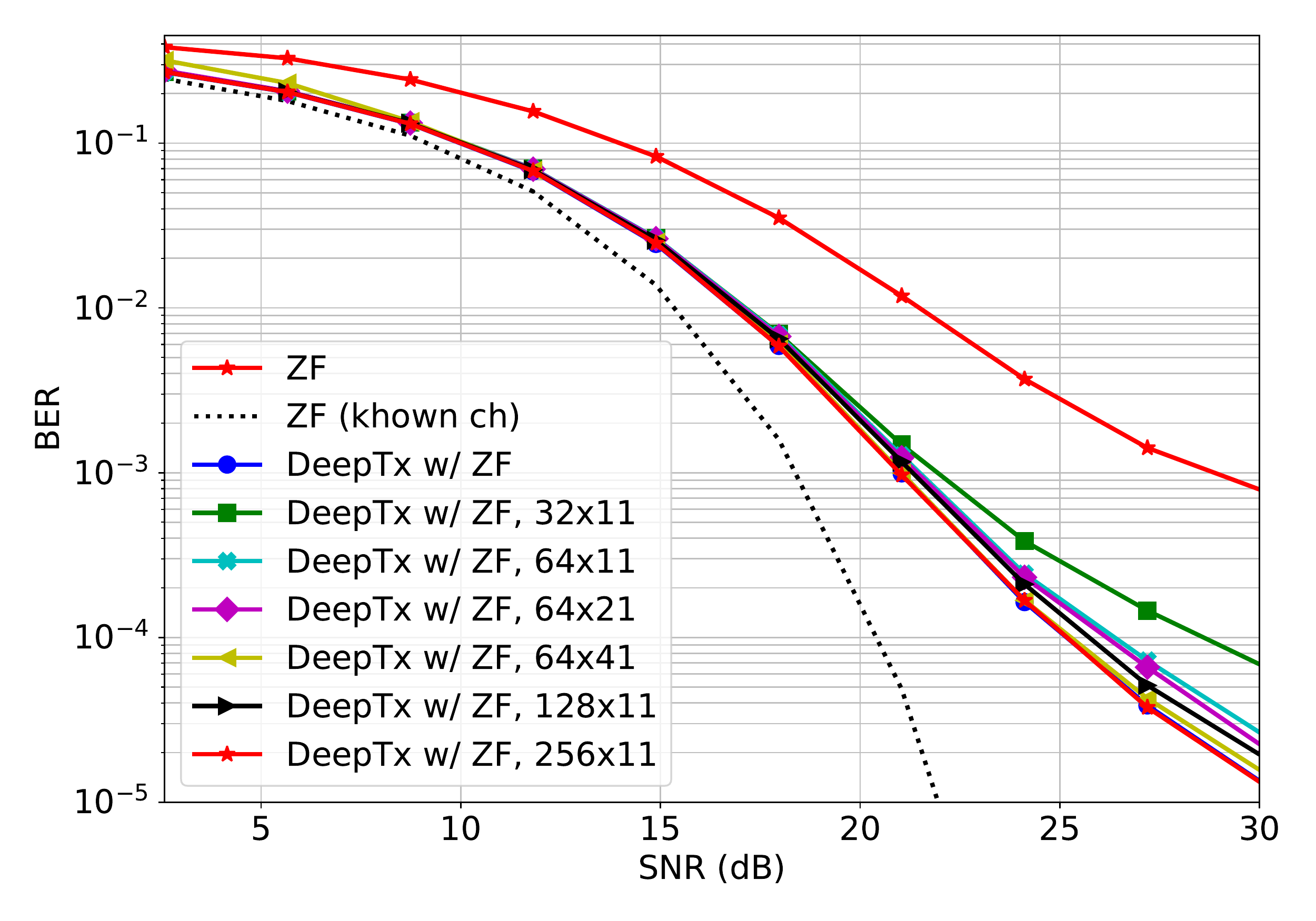} }}%
	~
	\subfloat[$\tau=6$]{{\includegraphics[width=0.51\columnwidth]{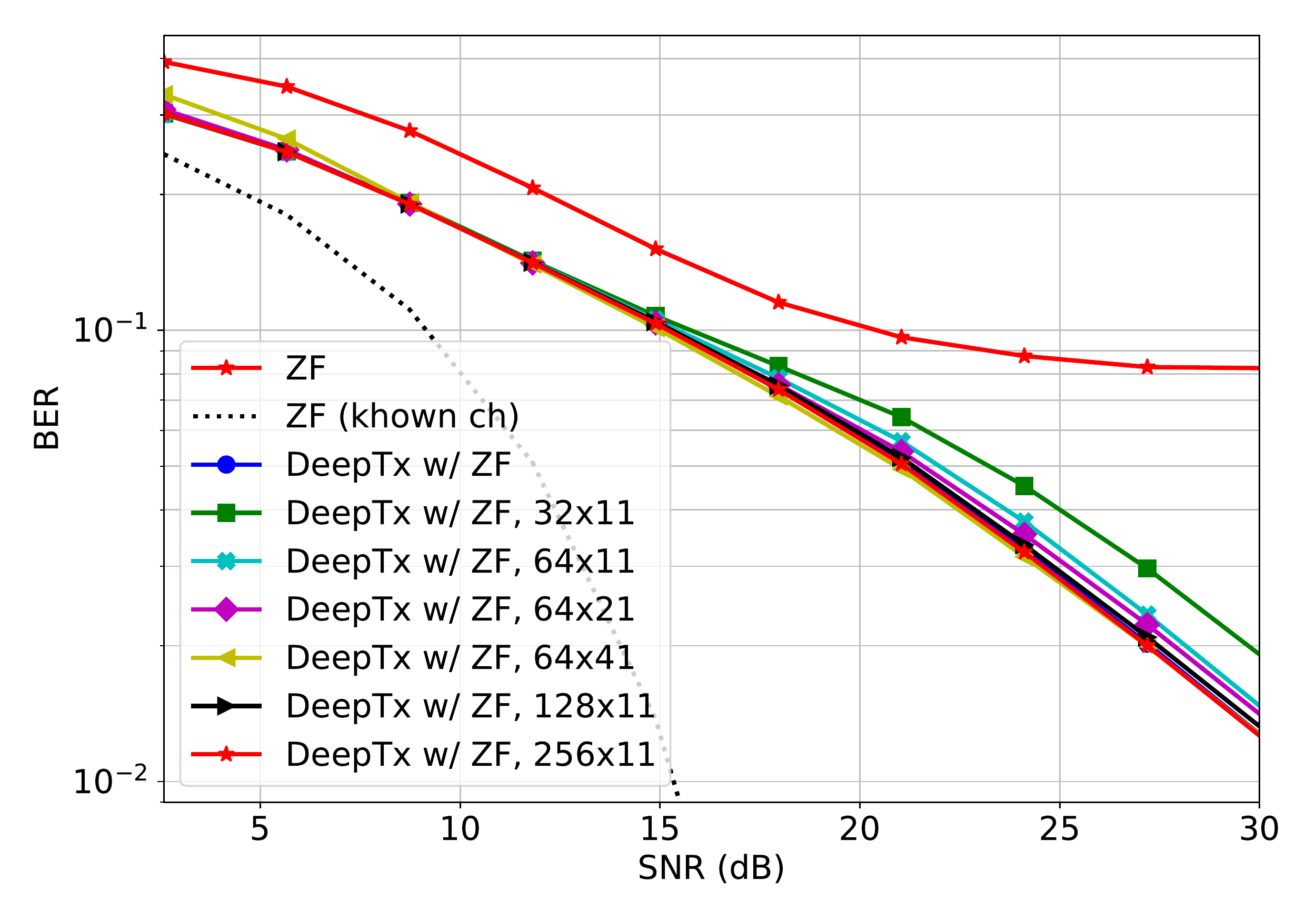} }}%
	\caption{Comparison of different model architectures for the prediction lengths $\tau=1$ (left) and $\tau=6$ (right).
        }%
	\label{fig:model_complexity}%
\end{figure*}

We also experimented with wide range of different ResNet architectures and hyperparameters to investigate what elements are essential for DeepTx performance. The networks are trained with the same approach as above, except for the altered model architecture and hyperparameters. The table reports average BER values for SNR range from 20~dB to 25~dB. For all runs, the optimizer hyperparameters were same as with the earlier validation results.  The results of this study are shown in Table \ref{tab:ablation}. A selection of results are also shown as BER curves in Figure \ref{fig:model_complexity}.

Results show that the original DeepTx architecture (Table \ref{tab:cnn_arch}) provides a good compromise in performance compared to experimented variations. The performance is slightly increased with wider model architectures, but these large models have also a significantly higher complexity. We also note that good alternatives can be achieved using a deeper model. For instance, 41 ResNet blocks with a channel width of 64 provides slightly better performance compared to the original model while having significantly fewer parameters. Furthermore, 21 ResNet blocks with 128 channels improves performance even more while having even fewer parameters. However, these models may not be optimal for hardware as increased depth may result in significantly higher latency.

We also observe that, as in DeepRx, the dilation is an essential feature in the model \cite{honkala21}. Performance is worse with shorter dilations, especially when no dilations are used. Dilations increase the receptive field of the CNN in an parameter-efficient manner and are most likely needed to model long-range dependencies in the channel over time and frequency.

\begin{table*}[!t]
	\renewcommand{\arraystretch}{1.3}
	\caption{Ablation study results (uncoded BER for uniformly distributed SNR between 20~dB and 25~dB). The architecture written in boldface refers to the primary DeepTx architecture (cf. Table \ref{table:param}). The labels XS, S, M, L and XL in the architecture name refer to extra small, small, medium, large and extra large in terms of number of parameters.}
	\label{tab:ablation}
	\centering
	\footnotesize
	\begin{tabular}{|l|p{1.4cm}|l|l|p{1.4cm}|p{1.5cm}|p{1.5cm}|p{1.5cm}|}
		\hline
		\textbf{Name} &   \textbf{Depth, Res-Net Blocks} & \textbf{Parameters} & \textbf{Channels} & \textbf{Min-max dilatation} &  \textbf{BER $\tau=1$} &   \textbf{BER $\tau=3$} & \textbf{BER $\tau=6$}  \\ \hline\hline
          \multicolumn{8}{|c|}{\emph{Variations of DeepTx architecture}} \\\hline
          XL & 13 & 4.8M  & 256-512 & 1-5  &  $4.31\cdot 10^{-4}$ &$7.03\cdot 10^{-3}$ &$4.05\cdot 10^{-2}$ \\\hline
L & 11 & 4.5M  & 256-512 & 1-5  &  $4.64\cdot 10^{-4}$ &$7.35\cdot 10^{-3}$ &$4.18\cdot 10^{-2}$ \\\hline
\textbf{DeepTx} & \textbf{11} & \textbf{963.1k}  & \textbf{128-256} & \textbf{1-5}  &  $\mathbf{5.09\cdot 10^{-4}}$ &$\mathbf{7.54\cdot 10^{-3}}$ &$\mathbf{4.21\cdot 10^{-2}}$ \\\hline
S & 9 & 387.3k  & 64-256 & 1-5  &  $6.55\cdot 10^{-4}$ &$8.78\cdot 10^{-3}$ &$4.69\cdot 10^{-2}$ \\\hline
XS & 9 & 66.4k  & 32-64 & 1-5  &  $7.68\cdot 10^{-4}$ &$1.03\cdot 10^{-2}$ &$5.14\cdot 10^{-2}$ \\\hline
XXS & 7 & 47.5k  & 32-64 & 1-5  &  $8.42\cdot 10^{-4}$ &$1.10\cdot 10^{-2}$ &$5.53\cdot 10^{-2}$ \\\hline
          \multicolumn{8}{|c|}{\emph{Constant widths, different depths}} \\\hline
32x11 & 11 & 34.8k  & 32 & 1-5  &  $8.59\cdot 10^{-4}$ &$1.12\cdot 10^{-2}$ &$5.48\cdot 10^{-2}$ \\\hline
64x11 & 11 & 114.7k  & 64 & 1-5  &  $6.78\cdot 10^{-4}$ &$8.88\cdot 10^{-3}$ &$4.72\cdot 10^{-2}$ \\\hline
64x21 & 21 & 209.4k  & 64 & 1-5  &  $6.54\cdot 10^{-4}$ &$8.31\cdot 10^{-3}$ &$4.44\cdot 10^{-2}$ \\\hline
64x41 & 41 & 398.9k  & 64 & 1-5  &  $5.08\cdot 10^{-4}$ &$7.18\cdot 10^{-3}$ &$4.02\cdot 10^{-2}$ \\\hline
128x11 & 11 & 409.6k  & 128 & 1-5  &  $6.11\cdot 10^{-4}$ &$7.90\cdot 10^{-3}$ &$4.28\cdot 10^{-2}$ \\\hline
256x11 & 11 & 1.5M  & 256 & 1-5  &  $5.02\cdot 10^{-4}$ &$7.32\cdot 10^{-3}$ &$4.13\cdot 10^{-2}$ \\\hline
256x21 & 21 & 2.9M  & 256 & 1-5  &  $4.26\cdot 10^{-4}$ &$6.69\cdot 10^{-3}$ &$3.86\cdot 10^{-2}$ \\\hline
\multicolumn{8}{|c|}{\emph{Smaller or no dilatation}} \\\hline
shorter dilatations & 11 & 963.1k  & 128-256 & 1-3  &  $5.75\cdot 10^{-4}$ &$8.35\cdot 10^{-3}$ &$4.54\cdot 10^{-2}$ \\\hline
no dilatations & 11 & 963.1k  & 128-256 & 1  &  $7.60\cdot 10^{-4}$ &$1.10\cdot 10^{-2}$ &$5.44\cdot 10^{-2}$ \\\hline
        \end{tabular}
\end{table*}

Scaling DeepTx to different scenarios may require models with larger complexity. For the architectures considered in this study, the asymptotic complexity for scaling to larger bandwidths and/or longer slots is $\mathcal{O}\left( S F \right)$ (i.e., linear with respect to subcarrier and time dimensions). Larger MIMO configurations with more antennas and MIMO layers will require larger models, for example, wider and deeper networks, but the relationship may not be linear with respect to the number of antennas and MIMO layers. Considering such scenarios is left as a future topic.

\section{Conclusion}
\label{sec:conc}

In this paper we considered machine learning based radio transmit beamforming in a time-division duplexing system. We combined a convolutional neural network (referred as DeepTx) with zero-forcing beamforming to create a mapping between a UL channel estimate and precoding coefficients in DL. The main task of DeepTx is to predict the evolution of the channel between UL slot and DL slots. In contrast to many previous works, it is trained in supervised manner by optimizing a loss function based on UE receiver performance. The training requires simulating both beamforming transmitter and UE receivers, as well as the channel between them, in a differentiable manner. 

The performance of DeepTx was evaluated using numerical simulations representing 5G data transmission. The results indicate that DeepTx significantly improves performance compared to traditional zero-forcing beamforming based on UL channel estimate.  To decrease computational burden, we also considered approximation of zero-forcing transformation in the approach using a series approximation. We found that relatively few terms in the approximation provides good results. 

We also showed that including information from several UL slots can significantly boost the performance, especially when UL channel estimates are based on a single pilot symbol in time (in which case a single slot cannot provide any information about evolution of channel). This approach, however, has a practical drawback as it requires lot of memory to store UL channel estimates for several UL slots and process them simultaneously. To overcome this, recurrent neural networks or long short term memory networks could potentially be applied for channel prediction in a recursive manner.

The study has also several limitations which should be addressed in future works. For example, we only considered a simple 4x2 MIMO setup, meaning that bigger MIMO systems should also be considered. However, this is left as a future topic since scaling the results to larger antenna arrays requires significantly more computational resources. Another limitation of the study is that we have only considered a case where the whole bandwidth is assigned to individual users. In practice, several users are often frequency-multiplexed on adjacent subcarriers. Historical channel information could also be utilized more efficiently, for example, with the help of RNN-based architectures.

Moreover, while we have experimentally studied several model architectures with a range of different computational complexities to investigate their radio performance, a further study on adapting DeepTx to a practical wireless setup is needed. However, we would like to emphasize that neural networks can be run with a very efficient inference time neural network chips, for example, with limited bit precision to reduce the latency and power usage. Therefore, a fully comprehensive complexity comparison against conventional receivers needs to target specific inference chip architectures. While such complexity study is outside the scope of this article, it constitutes an important future work item for us.

\section*{Acknowledgments}

This work has been partly funded by Business Finland (RAMPE project) and the European Commission through the H2020 project Hexa-X (Grant Agreement no. 101015956).



\begin{thebibliography}{10}
\providecommand{\url}[1]{#1}
\csname url@samestyle\endcsname
\providecommand{\newblock}{\relax}
\providecommand{\bibinfo}[2]{#2}
\providecommand{\BIBentrySTDinterwordspacing}{\spaceskip=0pt\relax}
\providecommand{\BIBentryALTinterwordstretchfactor}{4}
\providecommand{\BIBentryALTinterwordspacing}{\spaceskip=\fontdimen2\font plus
\BIBentryALTinterwordstretchfactor\fontdimen3\font minus
  \fontdimen4\font\relax}
\providecommand{\BIBforeignlanguage}[2]{{%
\expandafter\ifx\csname l@#1\endcsname\relax
\typeout{** WARNING: IEEEtran.bst: No hyphenation pattern has been}%
\typeout{** loaded for the language `#1'. Using the pattern for}%
\typeout{** the default language instead.}%
\else
\language=\csname l@#1\endcsname
\fi
#2}}
\providecommand{\BIBdecl}{\relax}
\BIBdecl

\bibitem{Zhang19a}
C.~{Zhang}, P.~{Patras}, and H.~{Haddadi}, ``Deep learning in mobile and
  wireless networking: A survey,'' \emph{IEEE Communications Surveys
  Tutorials}, vol.~21, no.~3, pp. 2224--2287, 2019.

\bibitem{Jiang17a}
C.~{Jiang}, H.~{Zhang}, Y.~{Ren}, Z.~{Han}, K.~{Chen}, and L.~{Hanzo},
  ``Machine learning paradigms for next-generation wireless networks,''
  \emph{IEEE Wireless Communications}, vol.~24, no.~2, pp. 98--105, 2017.

\bibitem{Fu18a}
Y.~{Fu}, S.~{Wang}, C.~{Wang}, X.~{Hong}, and S.~{McLaughlin}, ``Artificial
  intelligence to manage network traffic of {5G} wireless networks,''
  \emph{IEEE Network}, vol.~32, no.~6, pp. 58--64, 2018.

\bibitem{oshea17}
T.~{O'Shea} and J.~{Hoydis}, ``An introduction to deep learning for the
  physical layer,'' \emph{IEEE Transactions on Cognitive Communications and
  Networking}, vol.~3, no.~4, pp. 563--575, Dec 2017.

\bibitem{huang20}
H.~Huang, S.~Guo, G.~Gui, Z.~Yang, J.~Zhang, H.~Sari, and F.~Adachi, ``Deep
  learning for physical-layer {5G} wireless techniques: Opportunities,
  challenges and solutions,'' \emph{IEEE Wireless Communications}, vol.~27, no.
  214-222, 2020.

\bibitem{zhao2018}
Z.~Zhao, M.~C. Vuran, F.~Guo, and S.~Scott, ``Deep-waveform: A learned {OFDM}
  receiver based on deep complex-valued convolutional networks,'' \emph{IEEE
  Journal on Selected Areas in Communications}, vol.~39, no.~8, pp. 2407--2420,
  2021.

\bibitem{neumann18}
D.~{Neumann}, T.~{Wiese}, and W.~{Utschick}, ``Learning the {MMSE} channel
  estimator,'' \emph{IEEE Transactions on Signal Processing}, vol.~66, no.~11,
  pp. 2905--2917, June 2018.

\bibitem{chen19}
M.~Chen, U.~Challita, W.~Saad, C.~Yin, and M.~Debbah, ``Artificial neural
  networks-based machine learning for wireless networks: A tutorial,''
  \emph{IEEE Communications Surveys Tutorials}, vol.~21, no.~4, pp. 3039--3071,
  2019.

\bibitem{gunduz19}
D.~G\"und\"uz, P.~de~Kerret, N.~D. Sidiropoulos, D.~Gesbert, C.~R. Murthy, and
  M.~van~der Schaar, ``Machine learning in the air,'' \emph{IEEE Journal on
  Selected Areas in Communications}, vol.~37, no.~10, pp. 2184--2199, 2019.

\bibitem{honkala21}
M.~Honkala, D.~Korpi, and J.~M.~J. Huttunen, ``{DeepRx}: Fully convolutional
  deep learning receiver,'' \emph{IEEE Transactions on Wireless
  Communications}, vol.~20, no.~6, pp. 3925--3940, 2021.

\bibitem{korpi21}
D.~Korpi, M.~Honkala, J.~M.~J. Huttunen, and V.~Starck, ``{DeepRx} {MIMO}:
  Convolutional {MIMO} detection with learned multiplicative transformations,''
  in \emph{Proc. IEEE International Conference on Communications (ICC)}, 2021.

\bibitem{pihlajasalo21}
J.~Pihlajasalo, D.~Korpi, M.~Honkala, J.~M.~J. Huttunen, T.~Riihonen,
  J.~Talvitie, A.~Brihuega, M.~Uusitalo, and M.~Valkama, ``{HybridDeepRx}: Deep
  learning receiver for high-evm signals,'' in \emph{2021 IEEE 32nd Annual
  International Symposium on Personal, Indoor and Mobile Radio Communications
  (PIMRC),}, September 13-16, 2021.

\bibitem{korpi2022waveform}
D.~Korpi, M.~Honkala, J.~M.~J. Huttunen, F.~A. Aoudia, and J.~Hoydis,
  ``Waveform learning for reduced out-of-band emissions under a nonlinear power
  amplifier,'' arXiv:2201.05524, 2022.

\bibitem{goutay2021learning}
M.~Goutay, F.~A. Aoudia, J.~Hoydis, and J.-M. Gorce, ``End-to-end learning of
  ofdm waveforms with papr and aclr constraints,'' in \emph{2021 IEEE Globecom
  Workshops (GC Wkshps)}, 2021, pp. 1--6.

\bibitem{aoudia2021endtoend}
F.~A. Aoudia and J.~Hoydis, ``End-to-end waveform learning through joint
  optimization of pulse and constellation shaping,'' in \emph{Proc. IEEE
  Globecom Workshops}, 2021.

\bibitem{shental2019}
O.~Shental and J.~Hoydis, ``Machine {LLRning}: Learning to softly demodulate,''
  in \emph{Proc. IEEE Globecom Workshops}, 2019.

\bibitem{Bjornson2014}
E.~Bj\"ornson, M.~Bengtsson, and B.~Ottersten, ``Optimal multiuser transmit
  beamforming: A difficult problem with a simple solution structure [lecture
  notes],'' \emph{IEEE Signal Processing Magazine}, vol.~31, no.~4, pp.
  142--148, 2014.

\bibitem{zhang21}
J.~Zhang, Y.~Huang, J.~Wang, X.~You, and C.~Masouros, ``Intelligent interactive
  beam training for millimeter wave communications,'' \emph{IEEE Transactions
  on Wireless Communications}, vol.~20, no.~3, pp. 2034--2048, 2021.

\bibitem{kaya21}
A.~O. Kaya and H.~Viswanathan, ``Deep learning-based predictive beam management
  for {5G} {mmWave} systems,'' in \emph{Proc. IEEE Wireless Communications and
  Networking Conference (WCNC)}, 2021, pp. 1--7.

\bibitem{Xia2020}
W.~Xia, G.~Zheng, Y.~Zhu, J.~Zhang, J.~Wang, and A.~P. Petropulu, ``A deep
  learning framework for optimization of {MISO} downlink beamforming,''
  \emph{IEEE Transactions on Communications}, vol.~68, no.~3, pp. 1866--1880,
  2020.

\bibitem{Shi2020}
J.~Shi, W.~Wang, X.~Yi, X.~Gao, and G.~Y. Li, ``Deep learning-based robust
  precoding for massive mimo,'' \emph{IEEE Transactions on Communications},
  vol.~69, no.~11, pp. 7429--7443, 2021.

\bibitem{huang19}
H.~Huang, W.~Xia, J.~Xiong, J.~Yang, G.~Zheng, and X.~Zhu, ``Unsupervised
  learning-based fast beamforming design for downlink {MIMO},'' \emph{IEEE
  Access}, vol.~7, pp. 7599--7605, 2019.

\bibitem{Kong2021}
K.~Kong, W.-J. Song, and M.~Min, ``Knowledge distillation-aided end-to-end
  learning for linear precoding in multiuser {MIMO} downlink systems with
  finite-rate feedback,'' \emph{IEEE Transactions on Vehicular Technology},
  vol.~70, no.~10, pp. 11\,095--11\,100, 2021.

\bibitem{hu18}
Z.~Hu, J.~Cheng, Z.~Zhang, and Y.-C. Liang, ``Performance analysis of
  collaborative beamforming with outdated {CSI} for multi-relay spectrum
  sharing networks,'' \emph{IEEE Trans. Veh. Technol.}, vol.~67, no.~12, pp.
  11\,627--11\,641, 2018.

\bibitem{isukapalli09}
Y.~Isukapalli, R.~Annavajjala, and B.~D. Rao, ``Performance analysis of
  transmit beamforming for {MISO} systems with imperfect feedback,'' \emph{IEEE
  Trans. Commun.}, vol.~57, no. 222-231, 2009.

\bibitem{kim14}
J.-B. Kim, J.-W. Choi, and J.~M. Cioffi, ``Cooperative distributed beamforming
  with outdated {CSI} and channel estimation errors,'' \emph{IEEE Trans.
  Commun.}, vol.~62, no.~12, pp. 4269--4280, 2014.

\bibitem{eyceoz98}
T.~Eyceoz, A.~Duel-Hallen, and H.~Hallen, ``Deterministic channel modeling and
  long range prediction of fast fading mobile radio channels,'' \emph{IEEE
  Communications Letters}, vol.~2, no.~9, pp. 254--256, 1998.

\bibitem{arredondo02}
A.~Arredondo, K.~Dandekar, and G.~Xu, ``Vector channel modeling and prediction
  for the improvement of downlink received power,'' \emph{IEEE Transactions on
  Communications}, vol.~50, no.~7, pp. 1121--1129, 2002.

\bibitem{sternad03}
M.~Sternad and D.~Aronsson, ``Channel estimation and prediction for adaptive
  {OFDM} downlinks [vehicular applications],'' in \emph{Proc. IEEE 58th
  Vehicular Technology Conference (VTC 2003-Fall)}, vol.~2, 2003, pp.
  1283--1287 Vol.2.

\bibitem{jiang19}
W.~Jiang and H.~D. Schotten, ``A comparison of wireless channel predictors:
  Artificial intelligence versus {Kalman Filter},'' in \emph{ICC 2019 - 2019
  IEEE International Conference on Communications (ICC)}, 2019, pp. 1--6.

\bibitem{ding14}
T.~Ding and A.~Hirose, ``Fading channel prediction based on combination of
  complex-valued neural networks and chirp z-transform,'' \emph{IEEE Trans.
  Neural Netw. Learn. Syst.}, vol.~25, no.~9, pp. 1686--1695, 2014.

\bibitem{dong19}
P.~Dong, H.~Zhang, G.~Y. Li, I.~S. Gaspar, and N.~N. Alizadeh, ``Deep
  {CNN}-based channel estimation for {mmWave} massive {MIMO} systems,''
  \emph{IEEE Journal of Selected Topics in Signal Processing}, pp. 989--1000,
  2019.

\bibitem{potter08}
C.~Potter and A.~Panagos, ``{MIMO} channel prediction using recurrent neural
  networks,'' in \emph{Proc. Int. Telemetering Conf., San Diego, California},
  Oct. 2008.

\bibitem{routraya12}
G.~Routraya and P.~Kanungo, ``Genetic algorithm based {RNN} structure for
  rayleigh fading {MIMO} channel estimation,'' \emph{Procedia Engineering},
  vol.~30, pp. 77--84, 2012.

\bibitem{jiang18}
W.~Jiang and H.~D. Schotten, ``Neural network-based channel prediction and its
  performance in multi-antenna systems,'' in \emph{2018 IEEE 88th Vehicular
  Technology Conference (VTC-Fall)}, 2018, pp. 1--6.

\bibitem{jiang20}
------, ``Deep learning for fading channel prediction,'' \emph{IEEE Open
  Journal of the Communications Society}, vol.~1, pp. 320--332, 2020.

\bibitem{jiang20b}
W.~Jiang, M.~Strufe, and H.~Dieter~Schotten, ``Long-range {MIMO} channel
  prediction using recurrent neural networks,'' in \emph{2020 IEEE 17th Annual
  Consumer Communications Networking Conference (CCNC)}, 2020, pp. 1--6.

\bibitem{alrabeiah19}
M.~Alrabeiah and A.~Alkhateeb, ``Deep learning for {TDD} and {FDD} massive
  {MIMO}: Mapping channels in space and frequency,'' in \emph{Proc. 53rd
  Asilomar Conference on Signals, Systems, and Computers}, 2019, pp.
  1465--1470.

\bibitem{liao19}
Y.~Liao, Y.~Hua, X.~Dai, H.~Yao, and X.~Yang, ``Chanestnet: A deep learning
  based channel estimation for high-speed scenarios,'' in \emph{ICC 2019 - 2019
  IEEE International Conference on Communications (ICC)}, 2019, pp. 1--6.

\bibitem{luo20}
C.~Luo, J.~Ji, Q.~Wang, X.~Chen, and P.~Li, ``Channel state information
  prediction for {5G} wireless communications: A deep learning approach,''
  \emph{IEEE Transactions on Network Science and Engineering}, vol.~7, no.~1,
  pp. 227--236, 2020.

\bibitem{yuan19}
J.~Yuan, H.~Q. Ngo, and M.~Matthaiou, ``Machine learning-based channel
  estimation in {Massive MIMO} with channel aging,'' in \emph{2019 IEEE 20th
  International Workshop on Signal Processing Advances in Wireless
  Communications (SPAWC)}, 2019, pp. 1--5.

\bibitem{arnold2019}
M.~Arnold, S.~D\"orner, S.~Cammerer, J.~Hoydis, and S.~ten Brink, ``Towards
  practical {FDD} massive {MIMO}: {CSI} extrapolation driven by deep learning
  and actual channel measurements,'' in \emph{2019 53rd Asilomar Conference on
  Signals, Systems, and Computers}, 2019, pp. 1972--1976.

\bibitem{wang19}
J.~Wang, Y.~Ding, S.~Bian, Y.~Peng, M.~Liu, and G.~Gui, ``{UL-CSI} data driven
  deep learning for predicting {DL-CSI} in cellular {FDD} systems,'' \emph{IEEE
  Access}, vol.~7, pp. 96\,105--96\,112, 2019.

\bibitem{yang2020}
Y.~Yang, F.~Gao, Z.~Zhong, B.~Ai, and A.~Alkhateeb, ``Deep transfer learning
  based downlink channel prediction for {FDD} {Massive MIMO} systems,''
  \emph{IEEE Transactions on Communications}, vol.~68, no.~12, pp. 7485--7497,
  2020.

\bibitem{mashhadi21}
M.~B. Mashhadi and D.~G\"und\"uz, ``Pruning the pilots: Deep learning-based
  pilot design and channel estimation for {MIMO-OFDM} systems,'' \emph{IEEE
  Transactions on Wireless Communications}, vol.~20, no.~10, pp. 6315--6328,
  2021.

\bibitem{goutay21}
M.~Goutay, F.~A. Aoudia, J.~Hoydis, and J.-M. Gorce, ``Machine learning for
  {MU-MIMO} receive processing in {OFDM} systems,'' \emph{IEEE Journal on
  Selected Areas in Communications}, vol.~39, no.~8, pp. 2318--2332, 2021.

\bibitem{he2016identity}
K.~He, X.~Zhang, S.~Ren, and J.~Sun, ``Identity mappings in deep residual
  networks,'' in \emph{European conference on computer vision}.\hskip 1em plus
  0.5em minus 0.4em\relax Springer, 2016, pp. 630--645.

\bibitem{sifreM13}
L.~Sifre and S.~Mallat, ``Rotation, scaling and deformation invariant
  scattering for texture discrimination.'' in \emph{2013 IEEE Conference on
  Computer Vision and Pattern Recognition (CVPR)}.\hskip 1em plus 0.5em minus
  0.4em\relax IEEE Computer Society, 2013, pp. 1233--1240.

\bibitem{chollet2017}
F.~{Chollet}, ``Xception: Deep learning with depthwise separable
  convolutions,'' in \emph{2017 IEEE Conference on Computer Vision and Pattern
  Recognition (CVPR)}, 2017, pp. 1800--1807.

\bibitem{wu2013}
M.~Wu, B.~Yin, A.~Vosoughi, C.~Studer, J.~R. Cavallaro, and C.~Dick,
  ``Approximate matrix inversion for high-throughput data detection in the
  large-scale {MIMO} uplink,'' in \emph{2013 IEEE International Symposium on
  Circuits and Systems (ISCAS)}, 2013, pp. 2155--2158.

\bibitem{zhu15}
D.~Zhu, B.~Li, and P.~Liang, ``On the matrix inversion approximation based on
  neumann series in massive mimo systems,'' in \emph{2015 IEEE International
  Conference on Communications (ICC)}, 2015, pp. 1763--1769.

\bibitem{Matlab5G}
Mathworks, ``{Matlab 5G Toolbox},''
  \url{https://www.mathworks.com/products/5g.html}, 2020.

\bibitem{You2020Large}
Y.~You, J.~Li, S.~Reddi, J.~Hseu, S.~Kumar, S.~Bhojanapalli, X.~Song,
  J.~Demmel, K.~Keutzer, and C.-J. Hsieh, ``Large batch optimization for deep
  learning: Training {BERT} in 76 minutes,'' in \emph{International Conference
  on Learning Representations}, 2020.

\end{thebibliography}
\end{document}